\documentclass[11pt,twocolumn]{article}

\usepackage[margin=0.75in,top=0.5in,bottom=0.9in]{geometry}
\usepackage{amsmath, amssymb, graphicx, siunitx, xcolor, hyperref, subcaption}
\usepackage{textcomp}
\usepackage{url}
\usepackage{booktabs}
\usepackage{array}
\usepackage{cleveref}
\usepackage[version=4]{mhchem}
\usepackage[numbers]{natbib}
\usepackage{placeins}
\usepackage{setspace}
\usepackage{float}

\hypersetup{colorlinks=true, linkcolor=blue!60!black, citecolor=blue!60!black, urlcolor=blue!60!black}

\DeclareSIUnit\year{yr}

\makeatletter

\begin{document}
\twocolumn[
\begin{@twocolumnfalse}
\begin{center}
{\large \bfseries Self-subsidizing Mercury Remediation with Fusion Reactors\par}
\vspace{12pt}
{J.~F.~Parisi$^{1,\ast}$, J.~Azad$^{1}$\par}
\vspace{6pt}
{\small $^{1}$Marathon Fusion, 150 Mississippi Street, San Francisco, CA 94107, USA\\
$^\ast$e-mail: jason@marathonfusion.com\par}
\vspace{8pt}
\end{center}

\begin{quote}
\small
Fusion reactors can permanently remediate mercury by using it as a neutron multiplier: each (n,2n) reaction reduces the neutron number towards \ce{^{197}Hg}  which quickly decays into stable gold, irreversibly removing it from the environment while generating substantial economic value. Fusion energy is therefore not merely environmentally benign, but anti-polluting through the continuous consumption of an environmental pollutant. The history of nuclear fission demonstrates that environmental concerns can be decisive obstacles to low-carbon power deployment, suggesting that integrated pollution remediation fundamentally improves the policy calculus for fusion energy. We show that at high neutron flux (achievable in muon-catalyzed and inertial confinement fusion), nuclear reactions make all mercury isotopes eligible for gold transmutation, incentivizing mercury recovery and valuing the world mercury extractable stock at ${\sim}\$200$\,trillion, exceeding all in-ground gold reserves. Co-producing gold alongside electricity can triple a fusion plant's revenue, aligning economic incentives with complete, permanent mercury remediation.
\end{quote}
\vspace{12pt}
\end{@twocolumnfalse}
]
\makeatother
 
\section*{Introduction}

\begin{figure}[bt]
    \centering
    \includegraphics[width=\columnwidth]{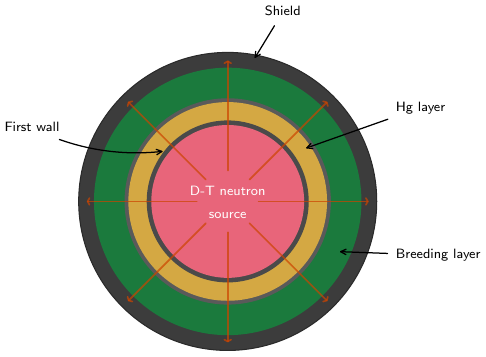}
    \caption{Schematic of a blanket geometry surrounding a D-T neutron source. Fusion neutrons radiate outward through the first wall into a liquid mercury layer, where successive (n,2n) reactions transmute Hg into Au. A tritium-breeding layer surrounds the Hg blanket to close the fuel cycle, enclosed by a radiation shield.}
    \label{fig:blanket_schematic}
\end{figure}

Mercury sits just one proton above gold in the periodic table. This proximity makes it well-suited to nuclear transmutation into stable gold with high-energy neutrons: a single neutron reaction strips one neutron from \ce{^{198}Hg}, producing \ce{^{197}Hg} via an (n,2n) reaction, which decays in days to stable \ce{^{197}Au}~\cite{Sherr1941}. Deuterium-tritium (D-T) fusion is an ideal driver for this transmutation at scale. Each D-T reaction releases a 14.1\,MeV neutron whose energy exceeds the (n,2n) threshold in mercury, and a gigawatt thermal plant produces ${\sim}3.6 \cdot 10^{20}$ neutrons per second. A fusion blanket (\cref{fig:blanket_schematic}) can sustain tritium self-sufficiency and transmute mercury into gold ~\cite{rutkowski2025scalable}. Because transmutation converts mercury into stable gold, it permanently removes mercury from the environment, offering a scalable remediation pathway for legacy stockpiles, industrial waste, and contaminated deposits~\cite{parsons2005brief,liu2012overview,budnik2019mercury}. The transmuted gold value strongly incentivizes environmental mercury recovery; for example, the feedstock value of the 300 to 900 tons of mercury released at Oak Ridge National Laboratory exceeds 10 times the estimated \$3.2 billion cleanup cost~\cite{GAO2024Mercury}. Co-producing gold alongside electricity in a fusion plant can also triple facility revenue at present gold prices~\cite{rutkowski2025scalable,parisi2025isotope}, reducing the levelized cost of fusion energy and potentially accelerating the path to commercial viability~\cite{Schwartz2023,Lindley2023}.

\begin{figure*}[t]
\centering
\includegraphics[width=2\columnwidth]{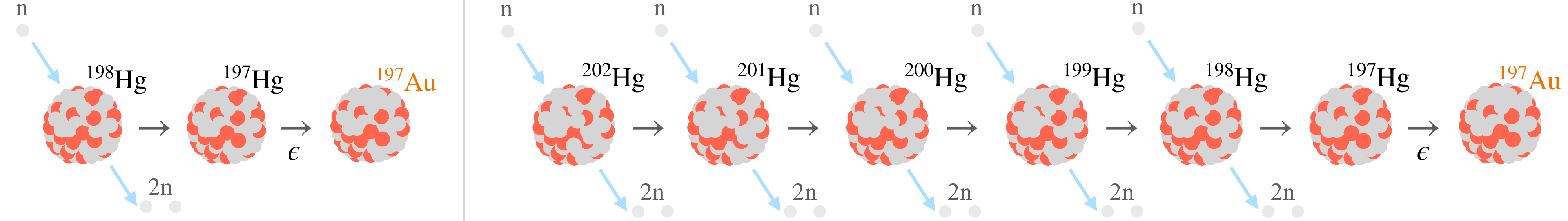}
\caption{Left: single-step chain of mercury transmutation into gold. Right: multi-step chain example.}
\label{fig:chains}
\end{figure*}

All prior analyses have treated \ce{^{198}Hg} as the sole productive isotope, with the heavier mercury isotopes (\ce{^{199}Hg} through \ce{^{204}Hg}) considered inert ballast~\cite{rutkowski2025scalable,parisi2025isotope}. This assumption is valid at magnetic-confinement-fusion (MCF) wall-loading fluxes ($\phi \sim 10^{14}$\,n\,cm$^{-2}$\,s$^{-1}$), where the characteristic transmutation time $\tau = 1/(\sigma\phi)$ exceeds the facility lifetime for multi-step chains. At the higher fluxes achievable in muon-catalyzed-fusion ($\mu$CF) ~\cite{Ponomarev1990,Froelich1992} systems ($\phi \sim 10^{15}$\,n\,cm$^{-2}$\,s$^{-1}$~\cite{parisi2025mucf}) and high-flux inertial-confinement-fusion (ICF) blankets ($\phi \gtrsim 10^{16}$\,n\,cm$^{-2}$\,s$^{-1}$)~\cite{meier2014,AbuShawareb2024}, $\tau$ drops to a few years and multi-step (n,2n) chains through stable intermediates become economically viable (\cref{fig:chains}). ICF blankets achieving high flux ($\phi \gtrsim 10^{16}$\,n\,cm$^{-2}$\,s$^{-1}$) may be possible by placing mercury feedstock much closer to the target~\cite{rutkowski2025scalable}. Higher flux also reduces the mercury blanket inventory, since the blanket mass optimistically scales as $M \propto 1/\phi$, from hundreds of tonnes at MCF flux to hundreds of kilograms at $\phi = 10^{17}$\,n\,cm$^{-2}$\,s$^{-1}$.

In this article we show that these multi-step chains make mercury one of the world's most valuable gold ores, and that transmutation permanently eliminates it as an environmental hazard. At $\phi = 10^{16}$\,n\,cm$^{-2}$\,s$^{-1}$, natural mercury is worth $\approx (2/3) p_\mathrm{Au}$ as gold feedstock with gold price $p_\mathrm{Au}$, with 87\% of this value arising from isotopes heavier than \ce{^{198}Hg}. The 1{,}500{,}000\,tonnes of ultimately recoverable mercury stock~\cite{Sverdrup2020} become worth ${\sim}\$174$\,trillion, exceeding all remaining in-ground gold reserves. While per-kilogram values converge at high neutron flux, a continuously replenished facility with 90\%at enriched \ce{^{198}Hg} feedstock produces ${\sim}3\times$ more gold, because a natural blanket must devote the majority of its neutron budget to an ``enrichment tax'': maintaining the blanket composition against dilution by fresh natural feed, rather than producing gold directly. Enrichment becomes increasingly cost-effective at higher flux, with returns of ${\sim}3\times$ at MCF to ${\sim}35\times$ at ICF. The multi-step chains ensure that all mercury, including enrichment tails, retains high transmutation value, guaranteeing near-complete resource utilization and remediation~\cite{Driscoll2013,Minamata2017}.

\section*{Results}

\textbf{Multi-step transmutation chains.}
Each stable mercury isotope with mass number $\mathrm{A} \geq 198$ can reach \ce{^{197}Au} via a sequence of (n,2n) reactions followed by electron capture
\begin{multline}
    \ce{^{A}Hg} \xrightarrow{(\mathrm{n,2n})} \ce{^{A-1}Hg} \xrightarrow{(\mathrm{n,2n})} \cdots \xrightarrow{(\mathrm{n,2n})} \ce{^{198}Hg} \\
    \xrightarrow{(\mathrm{n,2n})} \ce{^{197}Hg}/\ce{^{197\mathrm{m}}Hg} \xrightarrow{\varepsilon} \ce{^{197}Au},
    \label{eq:chain}
\end{multline}
requiring 
\begin{equation}
n_s \equiv \mathrm{A} - 197
\end{equation}
successive (n,2n) steps (\cref{fig:chains}). The single-step pathway from \ce{^{198}Hg} was first observed by Sherr, Bainbridge, and Anderson~\cite{Sherr1941} and recently proposed for fusion blankets~\cite{rutkowski2025scalable}. The ground state of \ce{^{197}Hg} undergoes electron capture to stable \ce{^{197}Au} with a half-life of 64.1\,h; the isomer \ce{^{197$\mathrm{m}$}Hg} ($T_{1/2} = 23.8$\,h) relaxes predominantly by isomeric transition (94.7\%) to the ground state, with a minor electron-capture branch (5.3\%) leading directly to gold~\cite{Kondev2021}. In either case, both species decay to stable gold.

A valuable feature of the mercury system is that all intermediates between most of the heavy stable isotopes (\ce{^{199}Hg} through \ce{^{202}Hg}) and \ce{^{198}Hg} are stable. The chain $\ce{^{202}Hg} \to \ce{^{201}Hg} \to \ce{^{200}Hg} \to \ce{^{199}Hg} \to \ce{^{198}Hg}$ passes through four stable intermediates with no decay bottleneck whatsoever. This contrasts sharply with other transmutation systems where radioactive intermediates create flux barriers requiring $\phi \gtrsim 10^{17}$ to $10^{18}$\,n\,cm$^{-2}$\,s$^{-1}$ to overcome. Only \ce{^{204}Hg} (6.8\% abundance) requires transit through radioactive \ce{^{203}Hg} ($t_{1/2} = 46.6$\,d), which becomes productive only above $\phi \gtrsim 10^{17}$\,n\,cm$^{-2}$\,s$^{-1}$. \Cref{tab:chains} summarizes the properties of stable mercury isotopes.

\begin{table}[bt]
\caption{Transmutation chains from stable Hg isotopes to \ce{^{197}Au}. Natural abundances from IUPAC 2021~\cite{iupac2021}. For isotopes 198 to 202, all intermediates are stable. Only \ce{^{204}Hg} must transit the radioactive \ce{^{203}Hg} ($t_{1/2} = 46.6$\,d).}
\label{tab:chains}
\centering
\begin{tabular}{lcc}
\toprule
Isotope & Natural abundance (\%) & (n,2n) steps \\
\midrule
\ce{^{196}Hg} & 0.15 & $-^\dagger$ \\
\ce{^{198}Hg} & 10.0 & 1 \\
\ce{^{199}Hg} & 16.9 & 2 \\
\ce{^{200}Hg} & 23.1 & 3 \\
\ce{^{201}Hg} & 13.2 & 4 \\
\ce{^{202}Hg} & 29.7 & 5 \\
\ce{^{204}Hg} & 6.8  & $7^\ddagger$ \\
\bottomrule
\end{tabular}
\vspace{4pt}
\small $^\dagger$\ce{^{196}Hg}(n,2n) makes \ce{^{195}Hg}/\ce{^{195}Au}, not \ce{^{197}Au}. However thermal-neutron capture can convert it to gold (see Supplementary Information). \\
\small $^\ddagger$Includes one radioactive intermediate (\ce{^{203}Hg}).
\end{table}

At 14.1\,MeV, the dominant neutron reaction on mercury isotopes is (n,2n), with cross sections of $\sigma \approx 2.0$\,b to $2.3$\,b across all isotopes~\cite{Temperley1969,AlAbyad2006,Shibata1997,Iwamoto2023,Brown2018} (Supplementary \cref{fig:cross_sections}). Competing channels such as (n,$\gamma$), (n,p), and (n,$\alpha$) have cross sections two to three orders of magnitude smaller at 14.1\,MeV, producing non-mercury, non-gold isotopes that exit the transmutation chain. We adopt a uniform effective cross section of $\sigma \approx 1.5$\,b, representing a flux-averaged value accounting for neutron moderation below the 14.1\,MeV birth energy (see Methods). At $\phi = 10^{16}$\,n\,cm$^{-2}$\,s$^{-1}$, the characteristic time per (n,2n) step is $\tau = 1/(\sigma\phi) \approx 2.1$\,yr. The time to peak \ce{^{197}Au} inventory for an $n_s$-step chain scales as $t_\mathrm{peak} \sim n_s\tau$: roughly 4\,yr for the 2-step \ce{^{199}Hg} chain, and 11\,yr for the 5-step \ce{^{202}Hg} chain. These timescales are comparable to the design lifetimes of fusion facilities, making the economics of multi-step chains viable at high flux.

\textbf{Flux-dependent isotope value.}
The key economic quantity is the net present value (NPV) of gold produced per kilogram of mercury isotope $A$ under continuous irradiation. Integrating the discounted gold production over time (see Methods and \cref{supp:derivation}), the NPV is
\begin{equation}
    V_\mathrm{A}(\phi) = \frac{197}{\mathrm{A}}\, p_\mathrm{Au}\, \mathcal{Y}_\mathrm{Au}(\phi) \prod_{i=1}^{n_s} \frac{\Gamma_i}{\rho + \Gamma_i + \lambda_i},
    \label{eq:VsubA}
\end{equation}
where $p_\mathrm{Au}$ is the gold price per kilogram, $\Gamma_i = \sigma_i \phi$ is the transmutation rate, $\sigma_i$ is the (n,2n) cross section and $\lambda_i$ the decay rate of the parent at step $i$ ($\lambda_i = 0$ for stable parents), and $\rho = \ln(1{+}r)$ is the continuous discount rate corresponding to annual rate $r$. The gold yield $\mathcal{Y}_\mathrm{Au}(\phi)$ accounts for the fact that the final step produces \ce{^{197}Hg}, which must decay to \ce{^{197}Au} before collection; at high flux, \ce{^{197}Hg} may be destroyed into \ce{^{196}Hg} by (n,2n), or less likely by other reactions such (n,$\gamma$) into \ce{^{198}Hg}, before decaying. This gold-precursor burnup defines a critical flux $\phi^\dag_\mathrm{Au} = \lambda_\mathrm{g}/\sigma \approx 2 \times 10^{18}$\,n\,cm$^{-2}$\,s$^{-1}$, above which more than half of the gold precursor is lost (see \cref{fig:gold_yield}). Each factor $\Gamma_i/(\Gamma_i + \lambda_i + \rho)$, captures both branching loss and the time cost of waiting. The factor $\Gamma_i/(\Gamma_i + \rho)$ per step can be rewritten as $1/(1 + \rho\tau)$, the present-value factor for a single exponentially distributed waiting time with mean $\tau = 1/\Gamma_i$. For chains through stable intermediates with uniform $\sigma_i$, the product simplifies to $[\Gamma_i/(\Gamma_i + \rho)]^{n_s}$. The isotope value $V_\mathrm{A}$ is therefore non-monotonic in flux, rising as multi-step chains become viable and declining when gold-precursor burnup dominates. $V_\mathrm{A}$ (\cref{eq:VsubA}) thus describes a competition between three rates: neutron-driven transmutation at $\Gamma_i$, the cost of money $\rho$, and (only for \ce{^{203}Hg}) decay $\lambda_i$.

\begin{figure*}[tb]
\centering
\begin{subfigure}[t]{0.44\textwidth}
\centering
\includegraphics[width=\textwidth]{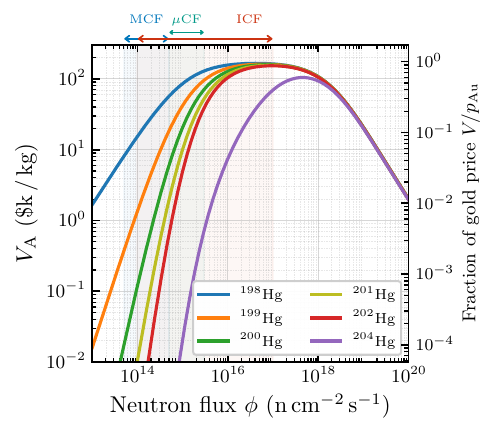}
\caption{}
\label{fig:isotope_value}
\end{subfigure}%
\hfill
\begin{subfigure}[t]{0.44\textwidth}
\centering
\includegraphics[width=\textwidth]{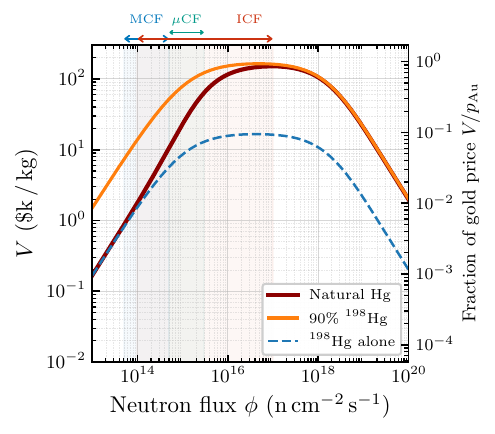}
\caption{}
\label{fig:natural_value}
\end{subfigure}%
\hfill
\begin{subfigure}[t]{0.44\textwidth}
\centering
\includegraphics[width=\textwidth]{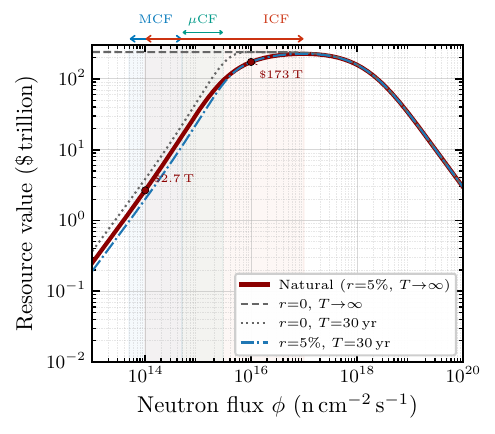}
\caption{}
\label{fig:reserve_value}
\end{subfigure}
\caption{(a) Gold-production NPV $V_\mathrm{A}$ per kilogram of each stable Hg isotope as a function of neutron flux ($p_\mathrm{Au} = \$175$k/kg, $r = 5\%$, $T \to \infty$). At MCF fluxes, only \ce{^{198}Hg} is productive; at ICF fluxes, all isotopes become valuable. (b) Value per kilogram of mercury for natural mercury (dark red), 90\%-enriched \ce{^{198}Hg} (orange), and the \ce{^{198}Hg} contribution alone (dashed blue). (c) Total transmutation value of world mercury extractable stock (1{,}500{,}000\,tonnes) under the same scenarios, plus reference curves showing effects of finite plant life and discounting separately.}
\label{fig:value_reserves}
\end{figure*}

\Cref{fig:isotope_value} shows $V_\mathrm{A}(\phi)$ for each isotope. At MCF fluxes ($\phi \sim 10^{14}$\,n\,cm$^{-2}$\,s$^{-1}$), only \ce{^{198}Hg} is productive. At ICF fluxes ($\phi \gtrsim 10^{16}$\,n\,cm$^{-2}$\,s$^{-1}$), all isotopes become valuable feedstocks. All curves turn over and decline above their optimal flux due to gold-precursor burnup, with the optimal flux scaling as $\phi_\mathrm{opt} \propto \sqrt{n_s}$ with chain length (see Methods).

Extra (n,2n) steps are cheap when transmutation is fast relative to the discount rate. At $\phi = 10^{16}$\,n\,cm$^{-2}$\,s$^{-1}$, the per-kilogram values converge. \ce{^{198}Hg} is worth $0.90\, p_\mathrm{Au}$ (at $r = 0.05$), while \ce{^{202}Hg}, despite requiring five chain steps, retains $0.59\, p_\mathrm{Au}$ (\Cref{tab:values}). This is far above the naive expectation that natural mercury (only 10\%at \ce{^{198}Hg}) should be worth ${\sim}1/10$ as much as pure \ce{^{198}Hg}. That reasoning treats the heavier isotopes as inert, but at high flux every isotope is a productive gold feedstock. The cost of extra chain steps is purely a discounting penalty: each step multiplies the value by $\Gamma_i/(\Gamma_i + \rho) \approx 0.91$ at $\phi = 10^{16}$\,n\,cm$^{-2}$\,s$^{-1}$ and $r = 0.05$. Five chain steps therefore reduce the per-kilogram value to $0.91^5 \approx 0.62$ of the single-step value.

\begin{figure}[tb]
\centering
\includegraphics[width=\columnwidth]{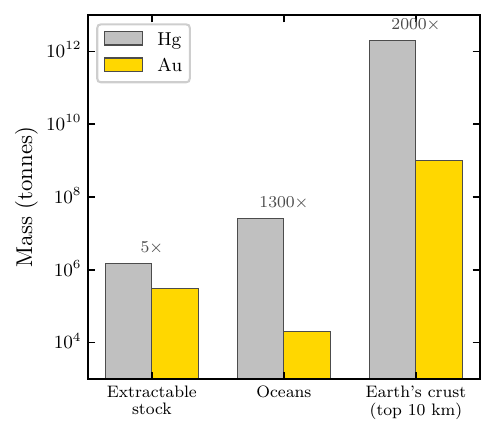}
\caption{Availability of mercury and gold at increasing levels of resource definition. ``Extractable stock'' uses Sverdrup and Olafsdottir~\cite{Sverdrup2020} for Hg (1{,}500{,}000\,t) and Sverdrup and Ragnarsdottir~\cite{Sverdrup2014} for Au (310{,}000\,t); ``Oceans'' are dissolved inventories~\cite{Lamborg2014}; ``Earth's crust'' is the total mass in the top 10\,km at average crustal abundances~\cite{Rudnick2014}. Mercury exceeds gold by 5 to 2{,}000$\times$ across all definitions.}
\label{fig:resource_hierarchy}
\end{figure}

\textbf{Natural mercury as gold ore.}
The value per kilogram of unenriched mercury is the abundance-weighted sum,
\begin{equation}
    V_\mathrm{nat} = \sum_\mathrm{A}  f_\mathrm{A} V_\mathrm{A} \simeq_{\phi = 10^{16}} 0.66\, p_\mathrm{Au},
    \label{eq:V_nat_value}
\end{equation}
where $f_\mathrm{A}$ is the mass fraction. We evaluated the sum at $\phi = 10^{16}$ n cm$^{-2}$ s$^{-1}$. Only ${\sim}\$15{,}500$/kg (14\%) comes from \ce{^{198}Hg}. At natural abundance, isotopes \ce{^{199}Hg} through \ce{^{202}Hg} collectively contribute 86\% of the total gold-production value at $\phi = 10^{16}$ n cm$^{-2}$ s$^{-1}$. One kilogram of natural mercury is worth $0.66\, p_\mathrm{Au}$ (${\sim}\$116{,}000$) as a gold feedstock, nearly four thousand times its commodity price of ${\sim}\$30$/kg~\cite{usgs2024,UNEP_Mercury2017}. Global mercury extractable stock of 1{,}500{,}000\,tonnes~\cite{Sverdrup2020} contains a gold-equivalent transmutation value of ${\sim}\$174$\,trillion (\cref{fig:reserve_value}), which exceeds the value of all remaining in-ground gold reserves (${\sim}59{,}000$\,tonnes, worth ${\sim}\$10$\,trillion at current prices~\cite{LBMA2026}). The commodity cost of acquiring this mercury extractable stock assuming current mercury prices held constant ($1{,}500{,}000$\,tonnes $\times$ \$30/kg $= \$45$\,B) is negligible compared to their gold-feedstock value. The mercury extractable stock contains at least an order of magnitude more gold than gold ore itself (\cref{fig:resource_hierarchy}).

This revaluation extends beyond conventional mercury mines: industrial sites, coal fly ash repositories, and legacy mining waste become economically valuable feedstock sources. At ICF flux, ${\sim}93\%$ of a natural mercury inventory converts to gold within 30\,yr (\Cref{fig:burnup_fraction}). The blanket mass falls as $M \propto 1/\phi$, from ${\sim}400$\,tonnes at MCF to ${\sim}4$\,tonnes at ICF flux (Table~2). The 1{,}500{,}000\,tonne extractable stock~\cite{Sverdrup2020} is likely a lower bound, since the ${\sim}3{,}900\times$ revaluation as gold feedstock would make additional mercury resources economic.

\textbf{Gold-production value per neutron.} We have discussed remediation from the perspective of NPV per unit of mercury. We now discuss the complementary approach of NPV per neutron produced from D-T reactions.

\begin{figure*}[tb]
\centering
\begin{subfigure}[t]{0.48\textwidth}
  \centering
  \includegraphics[width=\linewidth]{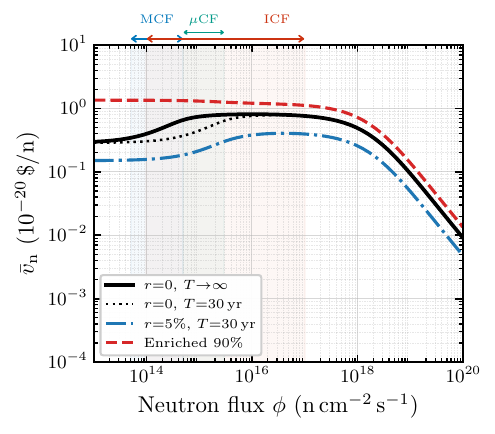}
  \caption{Gross gold-production NPV per source neutron $\bar{v}_\mathrm{n}$ (\Cref{eq:v_n_avg}), excluding mercury procurement costs. Natural mercury is shown at three $(r,T)$ scenarios: $r{=}0$, $T{\to}\infty$ gives the $v_0$ ceiling where every chain completes; finite $T$ suppresses long chains at low flux; finite $r$ adds a per-step discount penalty that vanishes at high flux. Enriched 90\% \ce{^{198}Hg} (red dashed) exceeds natural by ${\sim}10\times$ at MCF flux but converges to ${\sim}1.3\times$ at ICF flux as multi-step chains activate all isotopes.}
  \label{fig:neutron_value_a}
\end{subfigure}%
\hfill
\begin{subfigure}[t]{0.48\textwidth}
  \centering
  \includegraphics[width=\linewidth]{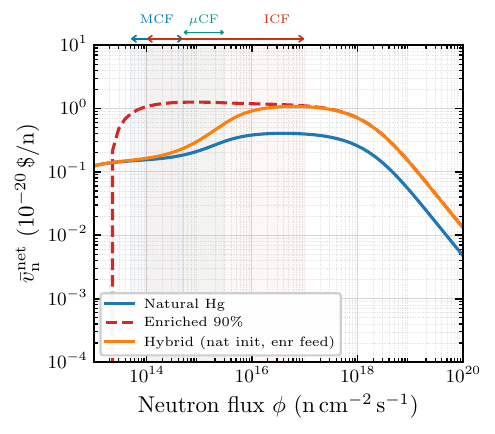}
  \caption{Net gold-production NPV per source neutron $\bar{v}_\mathrm{n}^\mathrm{net}$ (\Cref{eq:v_n_net}), after subtracting the initial blanket cost $M_0 c_\mathrm{init}$ and discounted feed cost ($r{=}5\%$, $T{=}30$\,yr). For natural mercury (blue), net $\approx$ gross because $c_\mathrm{init} \approx \$30$/kg is negligible. For enriched feedstock (red), the large initial blanket at MCF flux (${\sim}400$\,t at ${\sim}\$3{,}000$/kg) suppresses net well below gross; at ICF flux the blanket shrinks to ${\sim}4$\,t and the two converge. The hybrid strategy (orange) uses a natural initial loading and enriched feed.}
  \label{fig:neutron_value_b}
\end{subfigure}
\caption{Gold-production NPV per D-T source neutron ($\eta_\mathrm{pro} = 0.5$).}
\label{fig:neutron_value}
\end{figure*}

A direct measure of neutron productivity is the gold-production NPV per source neutron,
\begin{equation}
    \bar{v}_\mathrm{n}(\phi) = \frac{1}{\dot{N}\,T}\int_0^T \dot{m}_\mathrm{Au}(t)\,p_\mathrm{Au}\,e^{-\rho t}\,dt,
    \label{eq:v_n_avg}
\end{equation}
where $\dot{m}_\mathrm{Au}(t)$ is the gold production rate with continuous mercury replenishment~\cite{parisi2026neutronvalue}, $\dot{N}$ is the fusion neutron rate. In the limit $r = 0$, $T \to \infty$, all mercury is transmuted and $\bar{v}_\mathrm{n} \to v_0 \equiv \eta_\mathrm{pro}(197/N_A)\,p_\mathrm{Au}\,\mathcal{Y}_\mathrm{Au} \approx 2.9 \times 10^{-20}$\,\$/n, independent of composition or flux, because every neutron eventually leads to gold. Here, $\eta_\mathrm{pro}$ is the fraction of source D-T neutrons undergoing (n,2n) on mercury. Finite discounting penalizes each chain step by $\Gamma_i/(\Gamma_i + \rho)$, which vanishes at high flux ($\Gamma_i \gg \rho$); finite $T$ suppresses chains that cannot complete within the campaign. Equation~\eqref{eq:v_n_avg} is the gross revenue per neutron. The economically relevant quantity is the net NPV per neutron, which subtracts mercury capital costs,
\begin{equation}
\begin{aligned}
& \bar{v}_\mathrm{n}^\mathrm{net}(\phi) = \bar{v}_\mathrm{n}(\phi) - \\ 
& \frac{M_0 \left ( c_\mathrm{init} - c(E_\mathrm{T})\,e^{-\rho T} \right) + \displaystyle\int_0^T \dot{m}_\mathrm{Hg}(t)\,c_\mathrm{feed}\,e^{-\rho t}\,dt}{\dot{N}\,T},
\end{aligned}
\label{eq:v_n_net}
\end{equation}
where $M_0 = \eta_\mathrm{pro}\,\dot{N}\bar{A}/(\sigma\phi N_A)$ is the steady-state blanket mass, $\bar{A} \approx 200.6$\,g\,mol$^{-1}$ is the mean atomic mass of mercury, $\dot{m}_\mathrm{Hg}(t) = \dot{m}_\mathrm{Au}(t)\,(\bar{A}/197)$ is the mercury feed rate, $c_\mathrm{init}$ and $c_\mathrm{feed}$ are the per-kilogram costs of the initial blanket and replacement feed, and $c(E_\mathrm{T})$ is the per-kilogram value of the recovered blanket mercury at its terminal enrichment $E_\mathrm{T}$ (the \ce{^{198}Hg} fraction at $t = T$, which exceeds the feed enrichment due to in-situ transmutation of heavier isotopes). The $M_0\,c(E_\mathrm{T})\,e^{-\rho T}$ term is the discounted recovery value of the blanket at campaign end. Because $M_0 \propto 1/\phi$, both the initial cost and the terminal recovery scale as $1/\phi$, so their net present value $M_0[c_\mathrm{init} - c(E_\mathrm{T})\,e^{-\rho T}]$ is the dominant cost term for enriched feedstock at MCF flux, where blanket masses reach hundreds of tonnes; for natural mercury ($c_\mathrm{init} \approx \$30$/kg) the correction is negligible at all fluxes (\cref{fig:neutron_value_b}).

\Cref{fig:neutron_value_a} shows $\bar{v}_\mathrm{n}$ for natural mercury under three $(r, T)$ scenarios plus 90\%-enriched \ce{^{198}Hg}. At MCF flux, the natural blanket's NPV per neutron is ${\sim}10\times$ lower than enriched; at ICF flux the gap narrows to ${\sim}1.3\times$ as every neutron interaction creates comparable value. However, gold throughput does not converge: each gold atom from natural mercury consumes $\bar{n} \approx 3.6$ neutrons versus ${\sim}1$ for enriched. \cref{fig:neutron_value_b} shows the net NPV per neutron after subtracting mercury feed and blanket costs.

\textbf{Enrichment economics and the enrichment tax.}
Although \ce{^{198}Hg} has the highest per-kilogram value at any flux, enrichment does not increase the total gold-production NPV from a batch of natural mercury; it concentrates value into fewer kilograms of product while the tails retain the remainder. At $\phi = 10^{16}$\,n\,cm$^{-2}$\,s$^{-1}$, tails are worth ${\sim}0.63\, p_\mathrm{Au}$, so discarding them sacrifices ${\sim}85\%$ of the batch's gold value (\Cref{fig:value_fraction}). The per-kilogram marginal return from enrichment decreases with flux, from ${\sim}670\%$ at MCF to ${\sim}32\%$ at ICF, but this convergence is misleading since enrichment also increases gold production per neutron, and fusion neutron supply will likely be the binding constraint in the early days of fusion deployment.

\textbf{Neutron-scarce vs.\ mercury-scarce regimes.}
Whether neutron supply or mercury supply is the binding constraint determines the optimal strategy. When neutrons are scarce, enriched feed maximizes NPV per neutron, since a natural blanket diverts ${\sim}72\%$ of its neutron budget into isotope enrichment. When mercury is scarce, the conservation law (\cref{eq:conservation}) guarantees enrichment is neutral, so irradiating tails in a separate facility recovers the ${\sim}85\%$ of feed value otherwise lost. Early deployments likely favour enriched feed; at fleet scale, both regimes converge as all mercury is eventually consumed.

\begin{figure*}[tb]
\centering
\includegraphics[width=1.5\columnwidth]{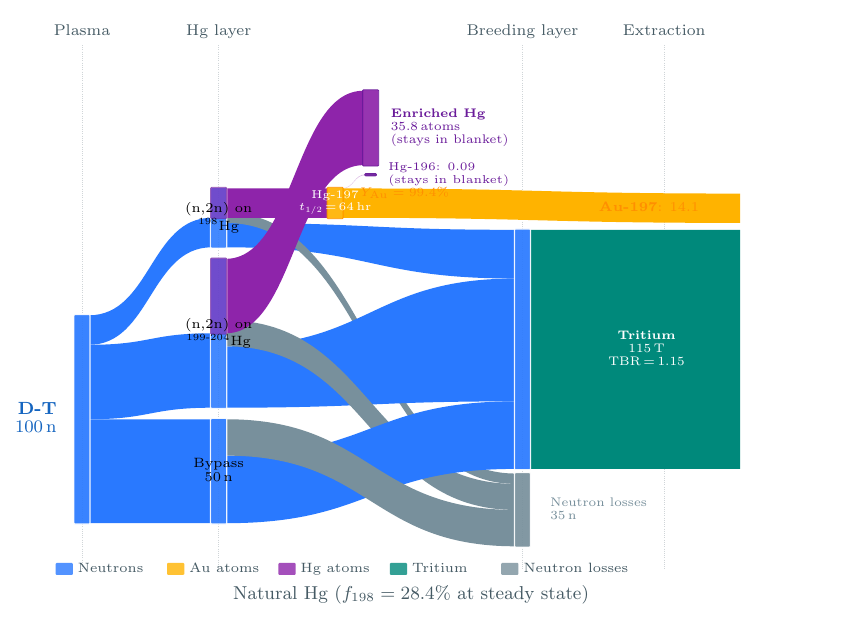}\\[1pt]
\includegraphics[width=1.5\columnwidth]{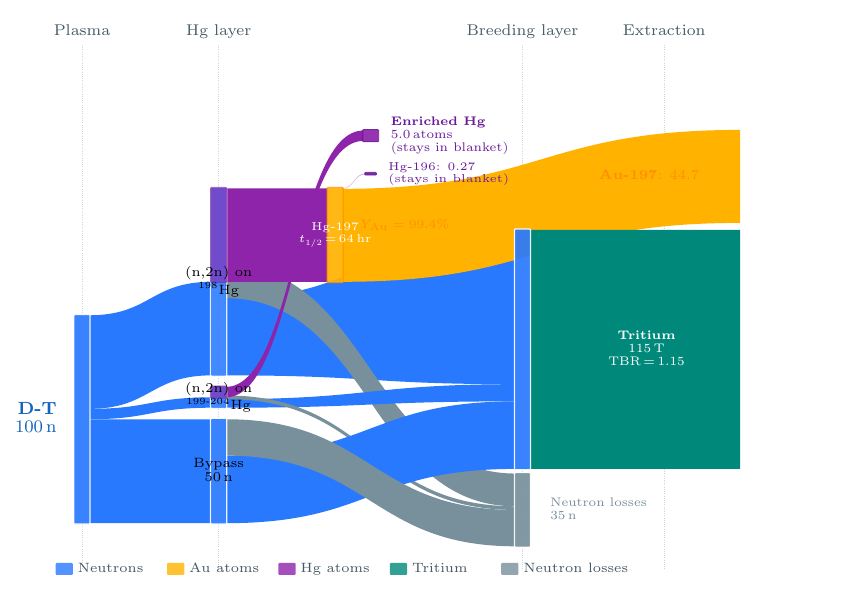}
\caption{Neutron and atom budget for a 1\,GW D-T facility at $\phi = 10^{16}$\,n\,cm$^{-2}$\,s$^{-1}$ in steady state ($\eta_\mathrm{pro} = 0.5$), comparing natural mercury (top) and 90\%-enriched \ce{^{198}Hg} (bottom). With natural mercury, only 28.4\% of the blanket is \ce{^{198}Hg} at steady state, so most reactions enrich heavier isotopes rather than producing gold (14.1 Au atoms per 100 neutrons). With enriched feedstock, 90\% of the blanket is \ce{^{198}Hg}, giving 44.7 Au atoms per 100 neutrons, a factor of ${\sim}3.2\times$ more.}
\label{fig:sankey}
\end{figure*}

In a natural mercury blanket at steady state, only 28.4\% of neutron interactions strike \ce{^{198}Hg} and produce gold directly; the remaining 71.6\% strike heavier isotopes, advancing them one step closer to gold but requiring the blanket to maintain its composition against perpetual dilution by fresh natural feed. We call this the \textit{enrichment tax}: $\sim$72\% of neutron interactions go to in-situ enrichment rather than gold extraction. With 90\%-enriched feed, the steady-state \ce{^{198}Hg} fraction rises to $79.2\%$ and the enrichment tax drops to $21\%$, so nearly all neutron value converts directly to gold revenue (\Cref{fig:sankey}).

This enrichment tax explains the persistent ${\sim}3\times$ throughput gap (\Cref{tab:facility}): the chain-length ratio $\bar{n}_\mathrm{nat}/\bar{n}_\mathrm{enr} \approx 3.6/1.3$ sets it. A hybrid strategy, starting with natural mercury and transitioning to enriched feed, captures ${\sim}95\%$ of the enrichment gain at $\phi = 10^{17}$\,n\,cm$^{-2}$\,s$^{-1}$ while avoiding the upfront blanket cost. The key parameter is the blanket turnover time $\tau_\mathrm{turn} = M/\dot{m}_\mathrm{feed} \propto 1/\phi$: at ICF flux ($\tau_\mathrm{turn} \approx 8$\,yr), feed composition dominates; at MCF flux ($\tau_\mathrm{turn} \sim 2{,}000$\,yr), initial loading matters most.

\textbf{Facility-level gold production.}
We compare representative facilities across three flux regimes (\Cref{tab:facility}). Consider a 1\,GW D-T source ($\dot{N} = 3.55 \times 10^{20}$\,n/s) with $\eta_\mathrm{pro} = 0.5$, continuously replenishing either natural mercury or 90\%-enriched \ce{^{198}Hg} over a 30\,yr campaign. The blanket mass follows $M = \eta_\mathrm{pro}\dot{N}\bar{A}/(\sigma\phi N_A)$, giving $M \propto 1/\phi$.

Three features stand out. First, the enriched facility consistently produces ${\sim}3\times$ more gold than natural, across all fluxes, because each gold atom from natural mercury consumes $\bar{n} \approx 3.6$ neutrons versus $\bar{n} \approx 1.3$ from enriched. This ratio is set by the mean chain length and is independent of flux. Second, the enrichment cost NPV drops from ${\sim}\$1.2$\,B at MCF flux (dominated by the 390\,t blanket) to ${\sim}\$70$\,M at ICF flux (dominated by the discounted 30\,yr feed); the return on enrichment rises from ${\sim}3\times$ to ${\sim}35\times$. Third, higher flux does not proportionally increase total gold production; the steady-state rate at $\phi = 10^{16}$\,n\,cm$^{-2}$\,s$^{-1}$ is actually ${\sim}10\%$ lower than at $\phi = 10^{14}$\,n\,cm$^{-2}$\,s$^{-1}$, mainly because heavier isotopes are actively cycling through chain intermediates. The (n,2n) reactions also constitute in-situ enrichment, driving the blanket composition toward lighter, more gold-proximate isotopes (\cref{fig:composition}): at $\phi = 10^{16}$\,n\,cm$^{-2}$\,s$^{-1}$, after 30\,yr, \ce{^{198}Hg} rises from 10\% to 28\% while heavier isotopes are partially consumed. At ICF flux, ${\sim}93\%$ of the mercury becomes gold within 30\,yr; at MCF flux, ${\sim}25\%$ of the value remains locked in partially converted mercury (\cref{fig:residual_value}(b)).

Gold-precursor burnup creates a flux optimum near $\phi_\mathrm{opt} \approx 10^{17}$\,n\,cm$^{-2}$\,s$^{-1}$, above which the gold yield $\mathcal{Y}_\mathrm{Au}$ collapses. A flow-through blanket eliminates this constraint by circulating mercury through the high-flux zone and decaying \ce{^{197}Hg} to gold outside the neutron field. With a duty cycle of $\tau_\mathrm{res}/\tau_\mathrm{cycle} = 10^{-3}$, the high-yield regime extends beyond $\phi = 10^{18}$\,n\,cm$^{-2}$\,s$^{-1}$ with $\mathcal{Y}_\mathrm{Au} > 99.9\%$ (\cref{fig:flow_through_yield}). OpenMC depletion simulations confirm the analytical model to within ${\sim}8\%$ under immediate-extraction conditions (\cref{supp:openmc}).

\begin{table}[tb]
\caption{Facility-level gold production from a 1\,GW${}_\mathrm{th}$ D-T source over 30\,yr ($\sigma = 1.5$\,b, $b = 0.5$, $\eta_\mathrm{pro} = 0.5$, $r = 5\%$). Nat: natural mercury ($\bar{n} = 3.6$); Enr: 90\%-enriched \ce{^{198}Hg} ($\bar{n} = 1.3$).}
\label{tab:facility}
\centering
\scriptsize
\setlength{\tabcolsep}{2pt}
\begin{tabular}{@{}lcccccc@{}}
\toprule
 & \multicolumn{2}{c}{$\phi{=}10^{14}$} & \multicolumn{2}{c}{$\phi{=}10^{16}$} & \multicolumn{2}{c}{$\phi{=}10^{17}$} \\
 & \multicolumn{2}{c}{(MCF)} & \multicolumn{2}{c}{(high-flux ICF)} & \multicolumn{2}{c}{} \\
\cmidrule(lr){2-3}\cmidrule(lr){4-5}\cmidrule(lr){6-7}
 & Nat & Enr & Nat & Enr & Nat & Enr \\
\midrule
$M_\mathrm{Hg}$ & 393\,t & 389\,t & 3.9\,t & 3.9\,t & 393\,kg & 389\,kg \\
Feed (kg/yr) & 320 & 1660 & 560 & 1470 & 480 & 1450 \\
Au mass (t) & 5.8 & 49 & 15 & 44 & 14 & 41 \\
Au NPV (\$B) & 0.53 & 4.5 & 1.3 & 4.1 & 1.3 & 3.8 \\
Enr.\ cost (\$M) & & 1250 & & 83 & & 70 \\
NPV gain (\$B) & & 4.0 & & 2.8 & & 2.5 \\
Return & & $3.2{\times}$ & & $34{\times}$ & & $36{\times}$ \\
\bottomrule
\end{tabular}
\end{table}

\section*{Discussion}

The 99.85\% of stable mercury isotopes with mass $A\geq197$ can be transmuted into stable gold in a fusion blanket while providing an essential function of neutron multiplication. This fundamentally transforms the economics of mercury remediation. The 90\% of natural mercury heavier than \ce{^{198}Hg}, previously considered inert, becomes productive at high neutron flux, with per-kilogram value reaching 86\% of the gold price at optimal flux. The case for natural mercury blankets is robust even though enriched feed produces ${\sim}3\times$ more gold per facility: enrichment infrastructure may be unavailable, mercury stockpiles are abundant, and multi-step chains guarantee that tails retain ${\sim}85\%$ of feed value, so no mercury is wasted regardless of processing strategy.

Several practical considerations support near-term feasibility. The conclusions above are robust to gold price uncertainty: even at the 10-year low, $V_\mathrm{nat} \approx \$26{,}000$/kg at high-flux ICF, far above the mercury commodity price, because all isotope values scale linearly with the gold price $p_\mathrm{Au}$. Mercury supply is also not a bottleneck. Each GW facility requires only 0.5 to 1.7\,tonnes/yr of feed (\cref{tab:facility}), modest relative to global mercury supply of ${\sim}3{,}500$ to $4{,}800$\,t/yr~\cite{UNEP_Mercury2017}. Existing government stockpiles (${\sim}4{,}400$\,t held by the US Defense Logistics Agency~\cite{USGS_MCS_Mercury2024}, plus European chlor-alkali surplus~\cite{EU_Reg_2017_852}) could supply early deployments without new mining, which the Minamata Convention prohibits~\cite{Minamata2017}. Importantly, mercury also serves as the primary neutron multiplier in the blanket, so once established as a component of the fusion fuel cycle, remediation incentives persist regardless of gold price fluctuations. The dominant quantitative uncertainty is the discount rate: varying $r$ from 0\% to 10\% changes $V_\mathrm{nat}$ from $1.0\, p_\mathrm{Au}$ to $0.41\, p_\mathrm{Au}$ at $\phi = 10^{16}$\,n\,cm$^{-2}$\,s$^{-1}$  (\cref{tab:sensitivity}).

These results reframe mercury as one of the most valuable stable elements, not for its chemical properties, but for its nuclear ones. A fusion plant co-producing electricity and gold \cite{rutkowski2025scalable} simultaneously addresses energy generation, precious-metal supply, and environmental remediation. At high neutron flux, 90 to 99\% of mercury converts to gold, and the revenue can defray the capital investment in fusion itself. Most importantly, transmutation is permanent. Unlike storage or containment, it eliminates mercury as a chemical element, leaving stable gold.

\section*{Methods}

Here we outline the main methods used in this work. More detailed derivations and further information is available in Supplementary Materials.

\textbf{Isotope value derivation.}
Consider a population $N_0$ of \ce{^{198}Hg} atoms placed in a constant neutron flux $\phi$ at $t = 0$. Each atom is transmuted by the (n,2n) reaction at rate $\Gamma = \sigma\phi$, producing \ce{^{197}Hg} which decays to \ce{^{197}Au}. The surviving population is $N(t) = N_0\, e^{-\Gamma t}$, and the NPV is the time integral of discounted gold revenue,
\begin{equation}
    V_{198} = \frac{197}{198}\, p_\mathrm{Au}\, \frac{\Gamma}{\Gamma + \rho},
    \label{eq:V_198}
\end{equation}
where $\rho = \ln(1{+}r)$ is the continuous discount rate. For an $n_s$-step chain starting from a single pure parent isotope, the Bateman chain equations~\cite{Bateman1910,Cetnar2006} can be solved by Laplace transform: the NPV equals $\Gamma_{n_s-1}\tilde{N}_{n_s-1}(\rho)$, where $\tilde{N}_{n_s-1}(\rho)$ is the Laplace transform of the penultimate-species population evaluated at $s = \rho$. This gives \Cref{eq:VsubA} (see Supplementary Information for the full derivation). For a target with arbitrary initial isotope composition $\{N_j(0)\}$, the same Laplace approach gives a general multi-isotope formula (Supplementary Information); setting $N_j(0) = f_j N_\mathrm{tot}$ recovers the abundance-weighted sum $V_\mathrm{nat} = \sum_A f_\mathrm{A} V_\mathrm{A}$.

\textbf{Gold-precursor burnup model.}
The final step produces \ce{^{197}Hg} or its isomer \ce{^{197m}Hg}, each facing three competing channels: decay to gold (rates $\lambda_g = \ln 2/(64.1\,\mathrm{h})$, $\lambda_m = \ln 2/(23.8\,\mathrm{h})$), (n,2n) burnup to \ce{^{196}Hg} (rate $\Gamma_{2n} = \sigma\phi$, a permanent loss), and radiative capture (n,$\gamma$) back to \ce{^{198}Hg} (rate $\Gamma_\gamma \approx 0.02\,\sigma\phi$, recycling for another attempt). The (n,$\gamma$) loss and recycling gain nearly cancel to leading order, as confirmed by our OpenMC validation. The self-consistent gold yield is $\mathcal{Y}_\mathrm{Au} = A_0/(1{-}\alpha)$ (see~\cref{supp:gold_yield} for more details), with $f_\varepsilon = 0.053$, $f_\mathrm{IT} = 0.947$, $b = 0.5$.

\textbf{Optimal flux.}
For an $n_s$-step chain, the $\Gamma$-dependent part of $\ln V_\mathrm{A}$ separates into chain benefit and burnup cost. Setting the derivative to zero in the regime $\rho \ll \Gamma \ll \lambda_{197}$ gives $\Gamma_\mathrm{opt}(n_s) = \sqrt{n_s\,\rho\,\lambda_{197}}$, or $\phi_\mathrm{opt}(n_s) = \Gamma_\mathrm{opt}/\sigma$. The $\sqrt{n_s}$ scaling arises because an $n_s$-step chain has $n_s$ times the incentive to raise the flux, balanced against the $n_s$-independent burnup cost.

\textbf{Finite irradiation time.}
For a blanket irradiated for time $T$, the NPV realized by campaign time $T$ is $V_\mathrm{A}(T, \phi) = V_\mathrm{A}^{\infty}(\phi)\; P(n_s,\, (\Gamma{+}\rho)\, T)$, where $P(n_s, x)$ is the regularized lower incomplete gamma function. At ICF flux ($\phi = 10^{16}$\,n\,cm$^{-2}$\,s$^{-1}$), 50\% of the asymptotic value is realized within ${\sim}5$\,yr and 90\% within ${\sim}11$\,yr.

\textbf{Neutron value.}
The blanket-average NPV per source neutron $\bar{v}_\mathrm{n}$ (\cref{eq:v_n_avg}) is computed from the blanket ODE with continuous replenishment~\cite{parisi2026neutronvalue}. The steady-state gold production rate is $\dot{m}_\mathrm{Au}^\mathrm{ss} = (\eta_\mathrm{pro}\dot{N}/\bar{n})\,(197/N_A)\,\mathcal{Y}_\mathrm{Au}$, where $\bar{n} = \sum_\mathrm{A} f_\mathrm{A}\, n_\mathrm{A}  \approx 3.6$ is the abundance-weighted mean chain length.

\textbf{Enrichment cost model.}
We estimate the enrichment cost from separative work unit (SWU) calculations~\cite{Benedict1981}. Mercury isotope enrichment has been demonstrated using centrifugal and electromagnetic methods~\cite{Love1973}. For $E = 0.9$, $c_\mathrm{Hg} \approx \$3{,}000$/kg.

\textbf{Cross sections.}
At 14.1\,MeV, (n,2n) cross sections for mercury cluster between 2.0 and 2.3\,b~\cite{Temperley1969,AlAbyad2006,Shibata1997,Iwamoto2023,Brown2018}. We adopt $\sigma \approx 1.5$\,b as a flux-averaged effective value. Using isotope-specific spectrum-averaged cross sections changes $V_\mathrm{nat}$ by at most $+6\%$ near $\phi \sim 10^{15}$ and $-10\%$ at $\phi \gtrsim 10^{18}$ (\Cref{tab:sensitivity}).

\textbf{Parameter values.}
Throughout this work: $p_\mathrm{Au} = \$175{,}000$/kg (2026)~\cite{LBMA2026,OConnor2015}, $r = 5\%$ ($\rho = 0.0488$\,yr$^{-1}$), $\sigma = 1.5$\,b, $b = 0.5$, $\eta_\mathrm{pro} = 0.5$. Natural abundances from IUPAC 2021~\cite{iupac2021}. Global mercury extractable stock of 1{,}500{,}000\,tonnes from Sverdrup and Olafsdottir~\cite{Sverdrup2020}.

\section*{Data availability}

Data for this study will be made available in a public repository upon publication.

\section*{Code availability}

Code used to generate all figures and perform the economic calculations will be made available in a public repository upon publication.

\section*{Acknowledgements}

We are grateful for discussions with A. Ahlholm, A. Rutkowski, and J. Wexler.

\bibliography{references}

\clearpage

\setcounter{figure}{0}
\setcounter{table}{0}
\setcounter{equation}{0}
\setcounter{section}{0}
\renewcommand{\thefigure}{S\arabic{figure}}
\renewcommand{\thetable}{S\arabic{table}}
\renewcommand{\theequation}{S\arabic{equation}}
\renewcommand{\thesection}{S\arabic{section}}

\makeatletter
\twocolumn[
\begin{@twocolumnfalse}
\begin{center}
{\LARGE\bfseries Supplementary Information\\[6pt]
\large Self-subsidizing Mercury Remediation with Fusion Reactors\par}
\vspace{8pt}
\end{center}
\end{@twocolumnfalse}
]
\makeatother

This Supplementary Information accompanies ``Self-subsidizing Mercury Remediation with Fusion Reactors.'' \Cref{supp:values} shows the gold-production NPV per kilogram for each stable mercury isotope at $\phi = 10^{16}$\,n\,cm$^{-2}$\,s$^{-1}$. \Cref{supp:derivation} derives the isotope value formula for single-step, multi-step Bateman-equation, and finite-time cases, and identifies the Erlang maturity factor as the fraction of asymptotic value realized by campaign time $T$. \Cref{supp:optimal_flux} derives the $\sqrt{n_s}$ scaling of the optimal neutron flux analytically from the competing discount and burnup penalties. \Cref{supp:sensitivity} quantifies how $V_\mathrm{nat}$ responds to uncertainty in the (n,2n) cross section, gold price, discount rate, and isomer branching ratio. \Cref{supp:openmc} validates the analytic blanket model against full OpenMC coupled depletion simulations and explains the role of continuous gold extraction. \Cref{supp:hg203} assesses the radiological burden from the steady-state \ce{^{203}Hg} inventory produced by (n,2n) on \ce{^{204}Hg}. \Cref{supp:intuition} provides intuition for the per-neutron value and the no-double-counting argument showing how value distributes across all neutrons in a chain. \Cref{supp:gold_yield} treats the competing (n,2n) and (n,$\gamma$) channels on the gold precursor \ce{^{197}Hg}, shows that their effects nearly cancel, and illustrates the full transmutation pathway on the chart of nuclides. \Cref{supp:full_conversion} describes two routes to near-complete mercury-to-gold conversion: thermal neutron capture on \ce{^{196}Hg} and a flow-through blanket design that decouples the transmutation and decay timescales. \Cref{supp:isotope_curves} shows the per-isotope value $V_\mathrm{A}(\phi)$ near the optimal-flux peak and the fractional contribution of each isotope to $V_\mathrm{nat}$. \Cref{supp:enrichment} derives the value-conservation law for enrichment, showing that separating \ce{^{198}Hg} merely redistributes gold-production value between product and tails, and quantifies the 85\% tails-value fraction at $\phi = 10^{16}$\,n\,cm$^{-2}$\,s$^{-1}$. \Cref{supp:blanket_economics} shows how the in-situ isotopic composition of a natural mercury blanket evolves over time and compares the cumulative discounted gold revenue for natural and enriched feeds. \Cref{supp:time_evolution} collects time-resolved profiles: blanket maturity curves, residual-value decomposition, and time-dependent gold production rates for two worked-example fluxes. \Cref{supp:conversion_sensitivity} shows the fraction of a natural mercury inventory converted to gold as a function of neutron flux and campaign duration, and the sensitivity of $V_\mathrm{nat}$ to the discount rate.

\section{Isotope value table}
\label{supp:values}

~\Cref{tab:values} lists the gold production value per kilogram of each stable mercury isotope at $\phi = 10^{16}$\,n\,cm$^{-2}$\,s$^{-1}$.

\begin{table}[h]
\caption{Gold-production value per kg of each stable Hg isotope at
$\phi = 10^{16}$\,n\,cm$^{-2}$\,s$^{-1}$ ($p_\mathrm{Au} = \$175$k/kg,
$r = 5\%$, $T \to \infty$, $\sigma \approx 1.5$\,b).
Columns: chain steps $n$; mean transmutation wait $n\tau$ (yr);
gold yield $\mathcal{Y}_\mathrm{Au}$; NPV per kg $V_\mathrm{A}$.}
\label{tab:values}
\centering\small
\begin{tabular}{lrrrr}
\toprule
Isotope & $n_s$ & $n_s\tau$ (yr) & $V_\mathrm{A}$ (\$/kg) & $V_\mathrm{A}/p_\mathrm{Au}$ \\
\midrule
\ce{^{198}Hg} & 1 &  2.1 & 156{,}900 & 0.90 \\
\ce{^{199}Hg} & 2 &  4.2 & 141{,}600 & 0.81 \\
\ce{^{200}Hg} & 3 &  6.3 & 127{,}700 & 0.73 \\
\ce{^{201}Hg} & 4 &  8.5 & 115{,}200 & 0.66 \\
\ce{^{202}Hg} & 5 & 10.6 & 103{,}900 & 0.59 \\
\ce{^{204}Hg} & 7 & 14.8 &   7{,}400 & 0.04 \\
\bottomrule
\end{tabular}
\end{table}

\begin{figure*}[tb]
\centering
\begin{subfigure}[t]{0.48\textwidth}
  \centering
  \includegraphics[width=\linewidth]{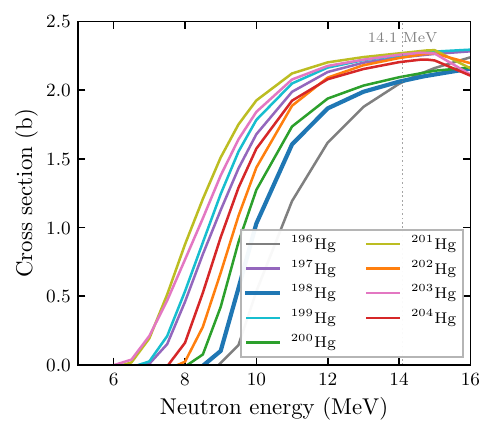}
  \caption{JENDL-5 (n,2n) cross sections for all stable and long-lived mercury isotopes. At 14.1\,MeV (dotted line), cross sections cluster between 2.0 and 2.3\,b.}
  \label{fig:cross_sections_a}
\end{subfigure}%
\hfill
\begin{subfigure}[t]{0.48\textwidth}
  \centering
  \includegraphics[width=\linewidth]{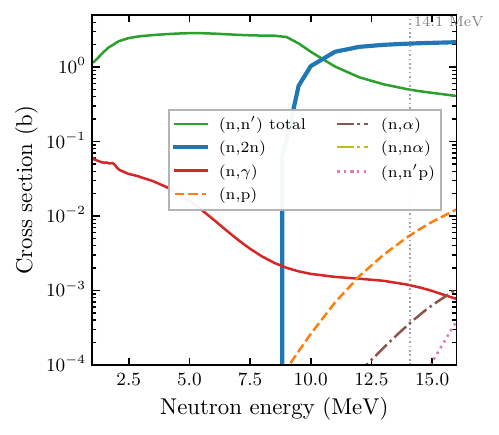}
  \caption{Competing reaction channels on \ce{^{198}Hg} from ENDF/B-VIII.0. At 14.1\,MeV, (n,2n) dominates by two to three orders of magnitude over (n,$\gamma$), (n,p), and (n,$\alpha$).}
  \label{fig:cross_sections_b}
\end{subfigure}
\caption{Neutron cross sections for mercury isotopes as a function of neutron energy.}
\label{fig:cross_sections}
\end{figure*}

\section{Derivation of the isotope value formula}
\label{supp:derivation}

We derive the net present value of gold produced per kilogram of isotope $A$ under continuous irradiation at constant flux $\phi$.

\subsection{Single-step chain ($n=1$)}

Consider a population $N_0$ of \ce{^{198}Hg} atoms placed in a constant neutron flux $\phi$ at $t = 0$. Each atom is transmuted by the (n,2n) reaction at rate $\Gamma = \sigma\phi$ (s$^{-1}$), producing \ce{^{197}Hg} which decays rapidly ($t_{1/2} = 64$\,h) to \ce{^{197}Au}. Here $\sigma$ (cm$^2$) is the effective (n,2n) cross section and $\phi$ (n\,cm$^{-2}$\,s$^{-1}$) is the neutron flux. The \ce{^{198}Hg} atom population evolves as
\begin{equation}
\dot{N} = - N \Gamma \,,
\end{equation}
with solution
\begin{equation}
    N(t) = N_0\, e^{-\Gamma t}.
\end{equation}
The gold production rate (in kg\,s$^{-1}$) is obtained by converting the atom depletion rate to mass. We assume each transmuted \ce{^{198}Hg} atom (mass $M_{198}/N_\mathrm{A}$) produces one \ce{^{197}Au} atom (mass $M_{197}/N_\mathrm{A}$), where $M_{198} = 198$\,g\,mol$^{-1}$ and $M_{197} = 197$\,g\,mol$^{-1}$ are the molar masses and $N_\mathrm{A}$ is Avogadro's number. Starting from initial mass $m_0$ (kg) of \ce{^{198}Hg}, the initial atom count is $N_0 = m_0 N_\mathrm{A}/M_{198}$, and the gold production rate is
\begin{align}
    \dot{m}_\mathrm{Au}(t)
        &= N_0\,\Gamma\,e^{-\Gamma t}\cdot\frac{M_{197}}{N_\mathrm{A}} \notag \\
        &= \frac{m_0 N_\mathrm{A}}{M_{198}}\,\Gamma\,e^{-\Gamma t}\cdot\frac{M_{197}}{N_\mathrm{A}} \notag \\
        &= \frac{197}{198}\,m_0\,\Gamma\,e^{-\Gamma t}.
    \label{eq:mdot_Au}
\end{align}
where the factor $197/198 = M_{197}/M_{198}$ is the dimensionless mass ratio converting consumed \ce{^{198}Hg} mass to produced \ce{^{197}Au} mass, and we have assumed \ce{^{197}Hg} decay is instantaneous relative to $1/\Gamma$.

\subsection{NPV integral}
\label{supp:npv}

Setting $m_0 = 1$\,kg, the revenue rate at time $t$ (in \$/s per kg of \ce{^{198}Hg}) is
\begin{equation}
    \text{revenue rate}(t) = \tfrac{197}{198}\,p_\mathrm{Au}\,\Gamma\,e^{-\Gamma t}.
\end{equation}
Gold produced at time $t$ is worth less today by the discount factor $e^{-\rho t}$, where
$\rho = \ln(1{+}r)$ is the continuous discount rate for annual rate $r$.
By compound interest, \$1 invested today grows to $(1+r)^t$ after $t$ years; equivalently,
\$1 received at time $t$ is worth $(1+r)^{-t} = e^{-\rho t}$ today.
The present value of revenue earned in the interval $[t,\,t+dt]$ is therefore
\begin{equation}
    dV = \tfrac{197}{198}\,p_\mathrm{Au}\,\Gamma\,e^{-\Gamma t}\,e^{-\rho t}\,dt.
\end{equation}
Integrating over all future time,
\begin{equation}
\begin{aligned}
V &= \tfrac{197}{198}\,p_\mathrm{Au}\,\Gamma \int_0^\infty e^{-(\Gamma+\rho)t}\,dt \\
& = \tfrac{197}{198}\,p_\mathrm{Au}\,\frac{\Gamma}{\Gamma+\rho} .\notag
\end{aligned}
\end{equation}
The factor $\Gamma/(\Gamma+\rho)$ is the fraction of gold production that occurs fast enough to retain economic value: when $\Gamma \gg \rho$ transmutation is fast and $V_{198} \to (197/198)\,p_\mathrm{Au}$, the full conversion value; when $\Gamma \ll \rho$ gold trickles out over centuries and the present value collapses to zero. Equivalently, writing $\tau = 1/\Gamma$ for the mean transmutation wait, $\Gamma/(\Gamma+\rho) = 1/(1+\rho\tau)$ is the present-value factor for a single exponentially distributed delay of mean $\tau$ discounted at rate $\rho$.

\begin{figure}[h]
    \centering
    \includegraphics[width=0.95\columnwidth]{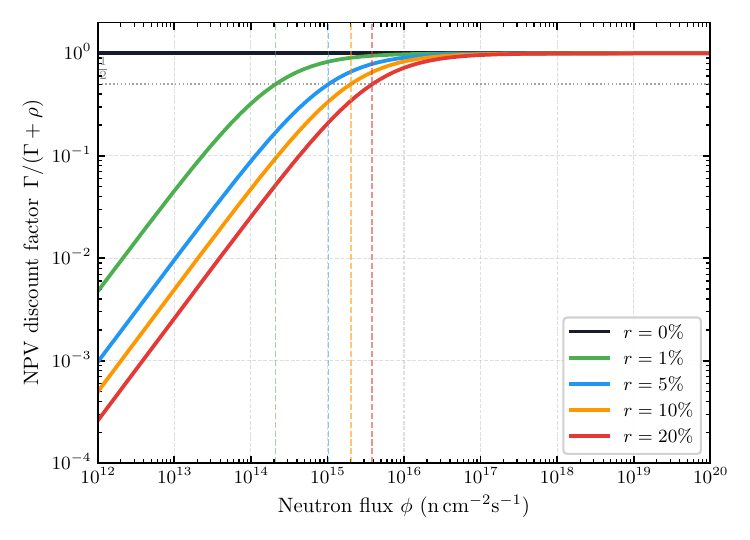}
    \caption{NPV discount factor $\Gamma/(\Gamma+\rho)$ versus neutron flux for $\sigma = 1.5$\,b and annual discount rates $r = 0\%$, $5\%$, $10\%$, and $20\%$. Vertical dashed lines mark the breakeven flux $\phi^* = \rho/\sigma$ for each non-zero rate; the dotted horizontal line marks the half-value point. Shaded bands indicate representative flux ranges for fission reactors and fusion blankets.}
    \label{fig:discount_factor}
\end{figure}

\Cref{fig:discount_factor} shows $\Gamma/(\Gamma+\rho)$ as a function of neutron flux for $\sigma = 1.5$\,b and four discount rates. The breakeven flux at which $\Gamma = \rho$ is $\phi^* = \rho/\sigma$; above $\phi^*$ most gold value is retained, below it the NPV is strongly suppressed. For $r = 5\%$ ($\rho \approx 0.049$\,yr$^{-1}$), $\phi^* \approx 10^{15}$\,n\,cm$^{-2}$\,s$^{-1}$, within the high-flux fusion-blanket regime.

\subsection{Multi-step chain through stable intermediates}
\label{supp:bateman}

For an $n$-step chain $\ce{^{A}Hg} \to \ce{^{A-1}Hg} \to \cdots \to \ce{^{198}Hg} \to \ce{^{197}Au}$ where all intermediates are stable and each step has rate $\Gamma_i = \sigma_i \phi$, the population dynamics form a Bateman decay chain~\cite{Bateman1910,Cetnar2006}. We allow arbitrary initial isotopic composition $\{N_k(0)\}$, which covers both enriched and natural mercury targets. The number density of the $k$-th species ($k = 0$ being the heaviest parent, $k = n$ being gold) satisfies, for $k \geq 1$,
\begin{equation}
    \dot{N}_k = \Gamma_{k-1} N_{k-1} - (\Gamma_k + \lambda_k)\, N_k,
\end{equation}
where $\lambda_k$ is the decay rate ($\lambda_k = 0$ for stable species). We use the property of the Laplace transform $\mathcal{L}\{\dot{N}_k\} = s\tilde{N}_k - N_k(0)$ where
\begin{equation}
\tilde{N}_k \equiv \mathcal{L}\{{N}_k\}.
\end{equation}
Rearranging gives
\begin{equation}
    \tilde{N}_k(s) = \frac{\Gamma_{k-1}}{s + \Gamma_k + \lambda_k}\, \tilde{N}_{k-1}(s)
                   + \frac{N_k(0)}{s + \Gamma_k + \lambda_k}, \quad k \geq 1.
\end{equation}
The second term carries the initial condition of species $k$ itself. Because the ODE is linear, contributions from each initial isotope superpose independently. Unrolling the recursion from position $j$ up to $n-1$, the contribution of species $j$ to $\tilde{N}_{n-1}(s)$ is
\begin{equation}
    \tilde{N}_{n-1}^{(j)}(s) = N_j(0)\;\frac{\displaystyle\prod_{k=j}^{n-2}\Gamma_k}
                                              {\displaystyle\prod_{k=j}^{n-1}(s+\Gamma_k+\lambda_k)},
    \label{eq:Ntilde_general}
\end{equation}
and the full solution is
\begin{equation}
\tilde{N}_{n-1}(s) = \sum_{j=0}^{n-1} \tilde{N}_{n-1}^{(j)}(s).
\end{equation}
The NPV is the discounted integral of the gold production rate $\Gamma_{n-1} N_{n-1}(t)$, i.e., the rate at which the last Hg isotope is transmuted to gold,
\begin{equation}
\begin{aligned}
V & = \frac{197\,p_\mathrm{Au}}{N_\mathrm{A}} \int_0^\infty \Gamma_{n-1}\,N_{n-1}(t)\,e^{-\rho t}\,dt \\
& = \frac{197\,p_\mathrm{Au}}{N_\mathrm{A}}\,\Gamma_{n-1}\,\tilde{N}_{n-1}(\rho).
\end{aligned}
\end{equation}
Substituting equation~\eqref{eq:Ntilde_general} at $s = \rho$, absorbing $\Gamma_{n-1}$ into the numerator product, and combining the two products over the same index range into one gives
\begin{equation}
    V = \frac{197\,p_\mathrm{Au}}{N_\mathrm{A}}\sum_{j=0}^{n-1} N_j(0)
        \prod_{k=j}^{n-1} \frac{\Gamma_k}{\rho+\Gamma_k+\lambda_k}.
    \label{eq:V_general}
\end{equation}
For a natural mercury target with $N_\mathrm{tot}$ atoms and isotopic abundance fractions $f_j$, substituting $N_j(0) = f_j N_\mathrm{tot}$ gives directly the NPV per atom of natural mercury as a weighted sum over isotopes, with each isotope contributing in proportion to its abundance and penalised by the number of transmutation steps it must traverse to reach gold.

For the special case of a pure single parent ($N_0(0) = N_\mathrm{tot}$, all others zero), only the $j=0$ term survives and equation~\eqref{eq:V_general} reduces to
\begin{equation}
    V_\mathrm{A} = \frac{197}{A}\, p_\mathrm{Au}\, \prod_{i=0}^{n-1} \frac{\Gamma_i}{\Gamma_i + \lambda_i + \rho}.
    \label{eq:V_A_derived}
\end{equation}
Each factor captures the competition between the transmutation rate $\Gamma_i$, any decay loss $\lambda_i$, and the economic time preference $\rho$. For a radioactive intermediate (e.g., \ce{^{203}Hg} in the \ce{^{204}Hg} chain), $\lambda_i > 0$ and the factor is suppressed unless $\Gamma_i \gg \lambda_i$, recovering the critical flux condition $\phi^\ast = \lambda/\sigma$. For stable intermediates, $\lambda_i = 0$ and each factor reduces to $\Gamma/(\Gamma + \rho)$.

In general, the NPV per kg of mercury with arbitrary isotopic distribution is
\begin{equation}
V = \sum_A \overline{f}_A V_\mathrm{A},
\end{equation}
where $\overline{f}_A$ is the fractional mass abundance of the mercury isotope A satisfying $\sum_A \overline{f}_A = 1$, and $V_{196} = 0$.

\subsection{Physical interpretation}

As shown in Section~\ref{supp:npv}, each factor $\Gamma_i/(\Gamma_i+\rho)$ is the present-value factor for a single exponentially distributed wait of mean $1/\Gamma_i$. The $n$-step product is therefore the present-value factor for $n$ sequential exponential waits, i.e., the compounded time-value penalty for traversing the full transmutation chain.

\subsection{Finite irradiation time}

For a blanket irradiated for time $T$, the gold production rate is proportional to the Erlang density $f_{n_s}(t) = \Gamma^{n_s} t^{{n_s}-1} e^{-\Gamma t}/({n_s}{-}1)!$, and the discounted integral becomes
\begin{align}
    V_\mathrm{A}(T) &= \frac{197}{A}\, p_\mathrm{Au}\, \frac{\Gamma^{n_s}}{({n_s}{-}1)!} \int_0^T t^{{n_s}-1} e^{-(\Gamma+\rho)t}\, dt \nonumber \\
    &= V_\mathrm{A}^\infty\; P\!\bigl({n_s},\, (\Gamma{+}\rho)\, T\bigr),
    \label{eq:V_time_derived}
\end{align}
where $P({n_s}, x)$ is the regularized lower incomplete gamma function (the CDF of the Erlang distribution with ${n_s}$ steps and rate $\alpha = \Gamma + \rho$). As $T \to \infty$, $P \to 1$ and the asymptotic result is recovered. The maturity fraction $P(n, (\Gamma{+}\rho)T)$ gives the fraction of asymptotic value realized by campaign time $T$.

\section{Derivation of optimal flux}
\label{supp:optimal_flux}

The $\Gamma$-dependent part of the log-value for an $n_s$-step chain separates into a ``chain benefit'' that grows with flux and a ``burnup cost'' that shrinks,
\begin{equation}
    \ln V_\mathrm{A} = \underbrace{n_s \ln\!\frac{\Gamma}{\Gamma + \rho}}_{\text{chain (discount)}} \;+\; \underbrace{\ln\!\frac{\lambda_{197}}{\lambda_{197} + \Gamma}}_{\text{burnup}} \;+\; \text{const.}
\end{equation}
Differentiating and setting $d(\ln V_\mathrm{A})/d\Gamma = 0$ gives
\begin{equation}
    \underbrace{\frac{n_s\,\rho}{\Gamma(\Gamma + \rho)}}_{\text{marginal chain benefit}} = \underbrace{\frac{1}{\lambda_{197} + \Gamma}}_{\text{marginal burnup cost}}.
\end{equation}
For the parameters of interest, three timescales are well separated; $\rho \approx 1.5 \times 10^{-9}$\,s$^{-1}$ $\ll$ $\Gamma$ $\ll$ $\lambda_{197} \approx 3 \times 10^{-6}$\,s$^{-1}$. In this regime the equation simplifies to
\begin{equation}
    \frac{n_s\,\rho}{\Gamma^2} = \frac{1}{\lambda_{197}}
    \quad\Longrightarrow\quad
    \Gamma_\mathrm{opt}(n_s) = \sqrt{n_s\,\rho\,\lambda_{197}},
\end{equation}
or $\phi_\mathrm{opt}(n_s) = \Gamma_\mathrm{opt}/\sigma$. An $n_s$-step chain has $n_s$~times the incentive to raise the flux compared to a single-step chain, giving the $\sqrt{n_s}$ scaling. For $n_s = 1$ (\ce{^{198}Hg}), $\phi_\mathrm{opt} \approx 4.5 \times 10^{16}$; for $n_s = 5$ (\ce{^{202}Hg}), $\phi_\mathrm{opt} \approx 1.0 \times 10^{17}$.

\section{Sensitivity and uncertainty}
\label{supp:sensitivity}

\subsection{Cross section}

The (n,2n) cross section $\sigma$ at 14.1\,MeV is the most important physical parameter. JENDL-5\cite{Iwamoto2023} and ENDF/B-VIII.0\cite{Brown2018} agree on $\sigma_{198} \approx 1.5 \pm 0.2$\,b, consistent with experimental measurements\cite{Temperley1969,AlAbyad2006}. The sensitivity of the isotope value is
\begin{equation}
    \frac{\partial \ln V_\mathrm{A}}{\partial \ln \sigma} = n_s \cdot \frac{\rho}{\Gamma + \rho}.
\end{equation}
At $\phi = 10^{16}$\,n\,cm$^{-2}$\,s$^{-1}$ ($\Gamma/\rho \approx 10$), even a 5-step chain shifts by only ${\sim}6\%$ for a $\pm 13\%$ change in $\sigma$. At MCF flux ($\phi = 10^{14}$\,n\,cm$^{-2}$\,s$^{-1}$), the sensitivity rises to ${\sim}0.9 n_s$.

The variation of $\sigma$ across isotopes introduces a systematic correction. Spectrum-averaged cross sections differ by $10$ to $17\%$ from the uniform-$\sigma$ assumption, producing a ${\sim}10\%$ enhancement in $V_\mathrm{nat}$ at ICF fluxes.

\subsection{Gold price}

All isotope values scale linearly with $p_\mathrm{Au}$. The 2026 spot price is \$175,000/kg; the 10-year range spans ${\sim}\$40{,}000$ to $\$175{,}000$/kg. At the 10-year low, $V_\mathrm{nat} \approx \$26{,}000$/kg at $\phi = 10^{16}$\,n\,cm$^{-2}$\,s$^{-1}$, far above the mercury commodity price. The qualitative conclusions hold across the full historical price range.

\subsection{Discount rate}

\begin{figure}[tb]
\centering
\includegraphics[width=\columnwidth]{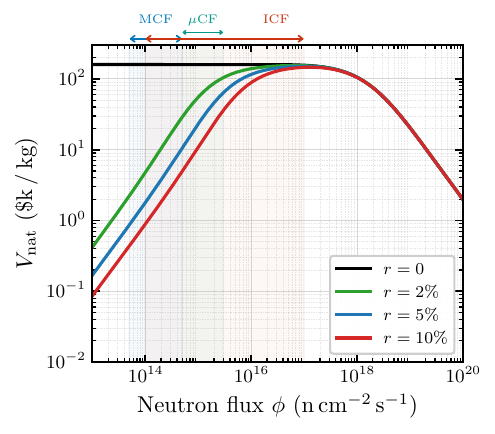}
\caption{Natural mercury value $V_\mathrm{nat}$ as a function of neutron flux for four discount rates ($T \to \infty$, $\sigma = 1.5$\,b). At $r = 0$, all chains complete and $V_\mathrm{nat} \to p_\mathrm{Au}$ below the burnup threshold. Finite $r$ suppresses value at low flux, with the penalty growing with chain length.}
\label{fig:discount_sensitivity}
\end{figure}

The discount rate $r$ sets the time cost of waiting for multi-step chains to complete (Supplementary \cref{fig:discount_sensitivity}). At $\phi = 10^{16}$\,n\,cm$^{-2}$\,s$^{-1}$, varying $r$ from 0\% to 10\% changes $V_\mathrm{nat}$ from $1.0\, p_\mathrm{Au}$ (\$175,000/kg) to $0.41\, p_\mathrm{Au}$ (\$72,000/kg). The choice of discount rate is the single largest source of quantitative uncertainty.

\subsection{Isomer branching ratio}

The fraction $b$ of (n,2n) reactions on \ce{^{198}Hg} that populate \ce{^{197$\mathrm{m}$}Hg} affects the gold yield $\mathcal{Y}_\mathrm{Au}$. We adopt $b = 0.5$; varying from $b = 0$ to $b = 1$ shifts $\mathcal{Y}_\mathrm{Au}$ by $< 1\%$ at $\phi \leq 10^{17}$\,n\,cm$^{-2}$\,s$^{-1}$ and by ${\sim}15\%$ at $\phi = 10^{18}$\,n\,cm$^{-2}$\,s$^{-1}$. This parameter matters only in the gold-burnup regime.

\subsection{Summary}

\Cref{tab:sensitivity} collects the sensitivities.

\begin{table}[h]
\caption{Sensitivity of $V_\mathrm{nat}$ at $\phi = 10^{16}$ to key input parameters.}
\label{tab:sensitivity}
\centering\small
\setlength{\tabcolsep}{4pt}
\begin{tabular}{lcc}
\toprule
Parameter & Variation & $\Delta V_\mathrm{nat}$ (\%) \\
\midrule
$\sigma$ & $\pm 13\%$ (1.3--1.7\,b) & $\pm 4$ \\
$p_\mathrm{Au}$ & $\pm 50\%$ & $\pm 50$ \\
$r$ & 0--10\% & $+51$/$-38$ \\
$b$ & 0--1 & $< 1$ \\
Per-isotope $\sigma_A$ & JENDL-5 & $+10$ \\
\bottomrule
\end{tabular}
\end{table}

\section{OpenMC validation}
\label{supp:openmc}

To validate the analytical blanket model against a full nuclear transport calculation, we performed coupled OpenMC depletion simulations\cite{Romano2015} using ENDF/B-VIII.0 cross sections\cite{Brown2018}.

\textit{Setup.} A thin slab of liquid mercury (0.1\,cm thick, 1\,cm$^2$ area) was irradiated with a monodirectional 14.1\,MeV neutron beam at $\phi = 10^{17}$\,n\,cm$^{-2}$\,s$^{-1}$ for 30\,yr (120 depletion steps of 0.25\,yr). At each step, all non-mercury products were extracted and fresh natural mercury added to maintain constant atom count.

\textit{Results.} The analytical ODE model ($\sigma = 1.5$\,b) captures the isotope dynamics correctly: \ce{^{198}Hg} enrichment from 10\% to ${\sim}24\%$, depletion of heavier isotopes, and approach to steady state within ${\sim}2$\,yr. At $t = 30$\,yr, the OpenMC and analytical \ce{^{198}Hg} fractions agree to ${\sim}1\%$ (0.246 vs.\ 0.249). The cumulative gold production from OpenMC (with Au burnup between extraction steps) is ${\sim}55\%$ of the analytical prediction. However, gold production inferred from the OpenMC \ce{^{198}Hg} time series assuming immediate extraction agrees with the analytical prediction to within ${\sim}8\%$, confirming the blanket ODE is accurate. The ${\sim}45\%$ shortfall arises from \ce{^{197}Au} burnup between extraction steps. At $\phi = 10^{17}$\,n\,cm$^{-2}$\,s$^{-1}$, the burnup half-life is ${\sim}37$\,days, so ${\sim}80\%$ of gold produced during each 0.25\,yr step is destroyed before extraction. This highlights the critical importance of rapid product extraction.

\begin{figure}[H]
\centering
\includegraphics[width=\columnwidth]{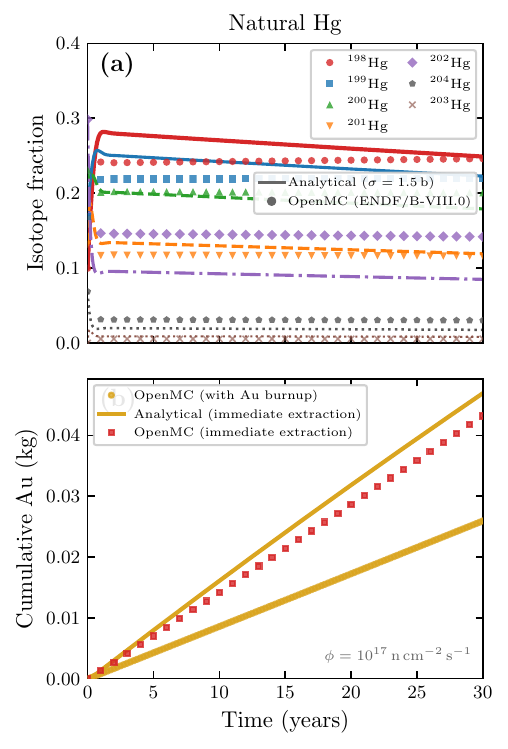}
\caption{Validation of the analytical blanket model against OpenMC coupled depletion at $\phi = 10^{17}$\,n\,cm$^{-2}$\,s$^{-1}$ with natural mercury feed. (a)~Time-dependent isotope fractions. (b)~Cumulative gold production. The close agreement between immediate-extraction inference and analytical prediction (${\sim}8\%$) validates the model; the shortfall in the actual OpenMC output arises from Au burnup between extraction steps.}
\label{fig:openmc_validation}
\end{figure}

\section{Radiological assessment of \texorpdfstring{\ce{^{203}Hg}}{Hg-203}}
\label{supp:hg203}

The (n,2n) reaction on \ce{^{204}Hg} (6.85\% of natural Hg) produces \ce{^{203}Hg} ($t_{1/2} = 46.6$\,d) at ${\sim}2{,}500\times$ the volumetric rate of \ce{^{203}Hg} from (n,$\alpha$) on \ce{^{206}Pb} in PbLi\cite{Merrill2014,Petti2006}. OpenMC depletion for a 1500\,MW plant\cite{rutkowski2025scalable} gives a steady-state inventory of 12.5\,kg (${\sim}6.4 \times 10^6$\,TBq). Despite being ${\sim}2{,}200\times$ less radiotoxic per Bq than \ce{^{210}Po} (derived air concentrations of $9.1 \times 10^3$ vs.\ $4.1$\,Bq/m$^3$\cite{ICRP68}), the DAC-equivalent burden is ${\sim}30\times$ that of \ce{^{210}Po} in PbLi. Mitigating factors include the absence of \ce{^{210}Po}, faster decay ($3\times$), and $\beta/\gamma$ rather than $\alpha$ emission; however, \ce{^{203}Hg} is ${\sim}10^4\times$ more volatile than PbPo\cite{Merrill2014}, requiring specialized safety measures. Selectively depleting \ce{^{204}Hg} to ${\sim}1\%$ would cut production by ${\sim}6\times$ with negligible impact on gold value.

\section{Intuition for the per-neutron value}
\label{supp:intuition}

The gold-production value per neutron striking isotope $A$ is
\begin{equation}
    v_\mathrm{n}(A, \phi) = \eta_\mathrm{pro}\,\frac{(A{-}1)}{N_A}\, V_{A-1}(\phi),
\end{equation}
where $V_{A-1}$ is the NPV per kilogram of the product isotope. For uniform cross sections, this simplifies to
\begin{equation}
    v_\mathrm{n}(A, \phi) = v_0(\phi)\, \left[\frac{\Gamma}{\Gamma + \rho}\right]^{\!A-198},
\end{equation}
where $v_0 \equiv \eta_\mathrm{pro}\,(197/N_A)\, p_\mathrm{Au}\, \mathcal{Y}_\mathrm{Au} \approx 2.9 \times 10^{-20}$\,\$/n is the single-step ceiling.

\textit{No discounting ($r = 0$, $T \to \infty$).} Every chain eventually completes, so $v_\mathrm{n}(A) = v_0$ for all isotopes. This is the flat ceiling in \cref{fig:neutron_value_a} of the main text.

\textit{Finite time ($r = 0$, $T = 30$\,yr).} The maturity factor $P(n, \Gamma T)$ reduces the value for chains too slow to complete in $T$, producing a curve that is flat at high flux but declines at low flux.

\textit{Finite discounting ($r = 5\%$).} The per-step factor $\Gamma/(\Gamma + \rho)$ penalizes each step. At ICF flux ($\Gamma/\rho \approx 100$), the penalty is negligible and $\bar{v}_\mathrm{n} \to v_0$. At MCF flux ($\Gamma/\rho \approx 0.1$), only \ce{^{198}Hg} contributes meaningfully.

There is no double-counting: the discount factors in $V_{A-1}$ already price in the expected cost of waiting for subsequent neutrons. No neutron is credited with more value than it creates; the value is distributed across all neutrons in the chain in proportion to their contribution to advancing mercury toward gold.

\section{Gold yield, burnup, and transmutation chains}
\label{supp:gold_yield}

\begin{figure*}[tb]
\centering
\includegraphics[width=1.4\columnwidth]{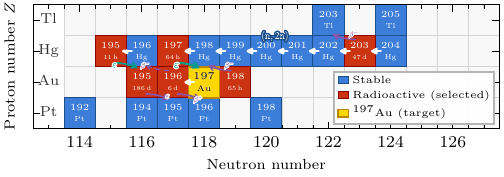}
\caption{Transmutation chains from stable mercury isotopes to \ce{^{197}Au} on the chart of nuclides. White arrows: successive (n,2n) reactions along the Hg chain. Green arrows: electron capture of \ce{^{197}Hg} and \ce{^{195}Hg} to gold. Purple arrows: $\beta^\pm$ and $\varepsilon$ decays of radioactive intermediates. Blue cells: stable nuclides; red cells: selected radioactive nuclides; gold cell: \ce{^{197}Au} product.}
\label{fig:chain_nz}
\end{figure*}

\subsection{Competing channels on the gold precursor}

Once \ce{^{198}Hg} undergoes (n,2n) to produce \ce{^{197}Hg} (or its isomer \ce{^{197m}Hg}), three processes compete for each precursor atom:
\begin{enumerate}
    \item \textbf{Decay to gold}: \ce{^{197}Hg} $\xrightarrow{\varepsilon}$ \ce{^{197}Au}, rate $\lambda_g = \ln 2/(64.1\,\mathrm{h})$; desired.
    \item \textbf{(n,2n) burnup}: \ce{^{197}Hg}(n,2n)\ce{^{196}Hg}, rate $\Gamma_{2n} = \sigma_{n,2n}^{(197)}\phi$; loss, since \ce{^{196}Hg} exits the gold chain.
    \item \textbf{(n,$\gamma$) recycling}: \ce{^{197}Hg}(n,$\gamma$)\ce{^{198}Hg}, rate $\Gamma_\gamma = \sigma_{n,\gamma}^{(197)}\phi$; the atom is sent \emph{back} to \ce{^{198}Hg} and gets a second chance to produce gold.
\end{enumerate}
The same three channels apply to \ce{^{197m}Hg}, with decay proceeding via isomeric transition (IT, fraction $f_{IT} = 0.947$, feeding the ground state) or direct electron capture (EC, fraction $f_\varepsilon = 0.053$, directly producing gold).

We adopt $\sigma_{n,2n}^{(197)} \approx 1.5$\,b (same as the chain steps) and estimate the spectrum-averaged \ce{^{197}Hg}(n,$\gamma$) cross section as $\sigma_{n,\gamma}^{(197)} \approx 30$\,mb. This estimate is based on (i) Hauser-Feshbach systematics for $A \approx 200$ nuclei at 14\,MeV giving $\sim$1--5\,mb for direct (n,$\gamma$); (ii) a moderated-neutron contribution from the blanket slowing-down spectrum; and (iii) the thermal value $\sigma_{n,\gamma}^{(197)} \approx 6.6$\,b~\cite{Mughabghab2006} providing an upper bound on the low-energy resonance contribution. This gives a ratio $R_{n\gamma} \equiv \sigma_{n,\gamma}^{(197)}/\sigma_{n,2n} \approx 30\,\mathrm{mb}/1500\,\mathrm{mb} = 0.02$. The uncertainty in $\sigma_{n,\gamma}^{(197)}$ is roughly a factor of three, corresponding to a range $R_{n\gamma} \in [0.007, 0.06]$.

\subsection{Self-consistent gold yield}

Because (n,$\gamma$) on \ce{^{197}Hg} recycles the atom to \ce{^{198}Hg}, the gold yield is a self-consistent quantity. Let $\mathcal{Y}_g$ and $\mathcal{Y}_m$ be the effective gold yields starting from one \ce{^{197}Hg} ground-state and isomer atom, respectively, including all recycling. Define total loss rates
\begin{equation}
    D_g = \lambda_g + \Gamma_{2n} + \Gamma_\gamma, \qquad D_m = \lambda_m + \Gamma_{2n} + \Gamma_\gamma,
\end{equation}
and let $Z = (1-b)\mathcal{Y}_g + b\mathcal{Y}_m$ be the gold yield per \ce{^{198}Hg} atom produced by recycling ($b = 0.5$ is the isomer branching fraction). The self-consistent equations are
\begin{equation}
    \mathcal{Y}_g = \frac{\lambda_g}{D_g} + \frac{\Gamma_\gamma}{D_g}\,Z, \qquad
    \mathcal{Y}_m = \frac{\lambda_m(f_\varepsilon + f_{IT}\mathcal{Y}_g)}{D_m} + \frac{\Gamma_\gamma}{D_m}\,Z.
\end{equation}
Substituting and collecting terms in $Z$ gives $Z = A_0/(1-\alpha)$, where
\begin{align}
    A_0 &= (1-b)\frac{\lambda_g}{D_g} + b\frac{\lambda_m}{D_m}\!\left(f_\varepsilon + f_{IT}\frac{\lambda_g}{D_g}\right), \label{eq:A0} \\
    \alpha &= (1-b)\frac{\Gamma_\gamma}{D_g} + b\frac{\Gamma_\gamma}{D_m}\!\left(1 + \frac{\lambda_m f_{IT}}{D_g}\right). \label{eq:alpha}
\end{align}
The effective gold yield per \ce{^{198}Hg} atom transmuted is therefore
\begin{equation}
    \mathcal{Y}_\mathrm{Au} = Z = \frac{A_0}{1-\alpha}.
    \label{eq:Y_Au_full}
\end{equation}
Here $A_0$ is the yield if (n,$\gamma$) acted purely as a loss (no recycling bonus), and $1/(1-\alpha)$ is the geometric series amplification from successive recycling events, with $\alpha$ the total recycling probability per cycle.

Setting $\Gamma_\gamma = 0$ recovers the previous model: $D_g \to \lambda_g + \Gamma_{2n}$, $\alpha \to 0$, and $A_0 \to \mathcal{Y}_\mathrm{Au}^{(0)}$.

\subsection{Near-cancellation of the two effects}

To leading order in $R_{n\gamma} = \Gamma_\gamma/\Gamma_{2n} \ll 1$, the loss from having $\Gamma_\gamma$ in $D_g$ and the recycling gain from $\alpha$ nearly cancel. At $\phi = 10^{16}$\,n\,cm$^{-2}$\,s$^{-1}$ ($\Gamma_{2n} \approx 470$\,yr$^{-1} \gg \lambda_g \approx 0.007$\,yr$^{-1}$), the correction to $\mathcal{Y}_\mathrm{Au}$ from the (n,$\gamma$) channel is $\mathcal{O}(R_{n\gamma}^2) < 0.05\%$. This near-cancellation occurs because every atom recycled via (n,$\gamma$) returns to \ce{^{198}Hg} and produces gold with essentially the same probability as the atom lost from direct decay. The net effect on $\mathcal{Y}_\mathrm{Au}$ is therefore negligible across the entire flux range of interest.

This confirms the agreement between our OpenMC depletion simulations and the analytic model (Supplementary \cref{fig:openmc_validation}): the (n,$\gamma$) channel on the gold precursor does not measurably affect gold yield, even though it is physically present. Equation~\eqref{eq:Y_Au_full} with $R_{n\gamma} = 0.02$ is used in all calculations.

\begin{figure}[H]
\centering
\includegraphics[width=\columnwidth]{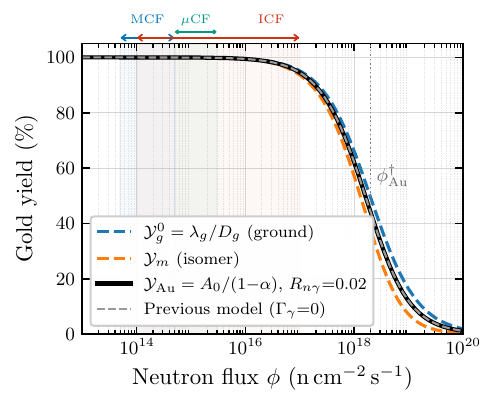}
\caption{Gold yield $\mathcal{Y}_\mathrm{Au}$ versus neutron flux for $b = 0.5$, including both (n,2n) and (n,$\gamma$) channels on \ce{^{197}Hg} with $\sigma_{n,\gamma}^{(197)} = 30$\,mb. The ground-state channel $\mathcal{Y}_g^0 = \lambda_g/D_g$ (blue dashed) drops to 50\% at the critical flux $\phi^\dag_\mathrm{Au} \approx 2\times10^{18}$\,n\,cm$^{-2}$\,s$^{-1}$. The isomer channel $\mathcal{Y}_m$ (orange dashed) falls more steeply. The effective yield $\mathcal{Y}_\mathrm{Au} = A_0/(1-\alpha)$ (solid black) includes the recycling correction; the dashed grey curve shows the previous model without (n,$\gamma$), confirming the near-cancellation.}
\label{fig:gold_yield}
\end{figure}

\section{Approaches to complete mercury-to-gold conversion}
\label{supp:full_conversion}

\begin{figure}[tb]
\centering
\includegraphics[width=\columnwidth]{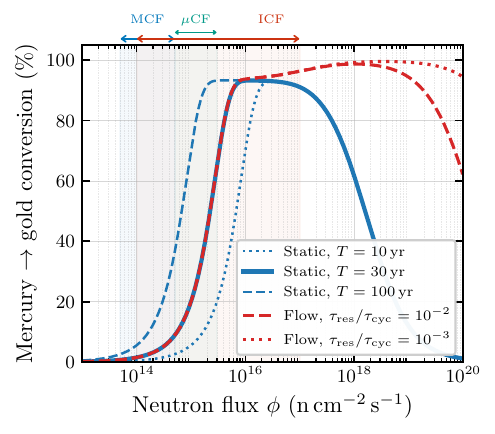}
\caption{Fraction of a natural mercury inventory converted to stable \ce{^{197}Au} as a function of neutron flux ($\sigma = 1.5$\,b, $b = 0.5$). Blue curves: static blanket for three campaign durations; conversion peaks near $\phi \sim 10^{16}$ and collapses at higher flux due to gold-precursor burnup. Red curves: flow-through blanket ($T = 30$\,yr) for two duty cycles $\tau_\mathrm{res}/\tau_\mathrm{cyc}$, which sustain near-complete conversion to arbitrarily high flux by decaying \ce{^{197}Hg} outside the neutron field.}
\label{fig:burnup_fraction}
\end{figure}

The gold conversion fraction saturates at ${\sim}93\%$ rather than unity (\cref{fig:burnup_fraction}). Three mechanisms limit the yield: (i) the $0.15\%$ that is \ce{^{196}Hg}, which cannot reach \ce{^{197}Au} via (n,2n); (ii) $\beta^-$ decay of \ce{^{203}Hg} ($t_{1/2} = 46.6$\,d) to stable \ce{^{203}Tl}, which diverts ${\sim}92\%$ of the \ce{^{204}Hg} chain at $\phi = 10^{16}$\,n\,cm$^{-2}$\,s$^{-1}$; and (iii) gold-precursor burnup ($\mathcal{Y}_\mathrm{Au} < 1$), which becomes significant above $\phi \gtrsim 10^{17}$\,n\,cm$^{-2}$\,s$^{-1}$.

\subsection{\ce{^{196}Hg} via thermal neutron capture}

Unlike heavier isotopes, \ce{^{196}Hg} can reach gold through radiative capture: \ce{^{196}Hg}(n,$\gamma$)\ce{^{197}Hg} $\to$ \ce{^{197}Au}. At thermal energies, $\sigma_\gamma(\ce{^{196}Hg}) \approx 3{,}080$\,b\cite{Mughabghab2006}, three orders of magnitude above the 14\,MeV (n,2n) cross section. The parasitic capture $\sigma_\gamma(\ce{^{197}Hg}) \approx 6.6$\,b is ${\sim}470\times$ smaller, ensuring efficient conversion. At $\phi_\mathrm{th} = 10^{14}$\,n\,cm$^{-2}$\,s$^{-1}$, $>99\%$ of \ce{^{196}Hg} converts within ${\sim}6$ months with negligible parasitic capture. The irradiation spectrum must be well thermalized, with negligible flux above ${\sim}1$\,eV.

\subsection{Flow-through blanket design}

Gold-precursor burnup can be circumvented by a flow-through design where liquid mercury circulates through the high-flux zone with residence time $\tau_\mathrm{res}$, then passes to an external hold-up volume where \ce{^{197}Hg} decays to gold outside the neutron field. The effective gold yield is
\begin{equation}
    \mathcal{Y}_\mathrm{Au}^\mathrm{flow} \approx \frac{\lambda_\mathrm{g}}{\lambda_\mathrm{g} + \sigma\phi\,(\tau_\mathrm{res}/\tau_\mathrm{cycle})},
\end{equation}
where $\tau_\mathrm{cycle} = \tau_\mathrm{res} + \tau_\mathrm{hold}$ is the total loop period. In the limit $\tau_\mathrm{res}/\tau_\mathrm{cycle} \to 0$, $\mathcal{Y}_\mathrm{Au}^\mathrm{flow} \to 1$ regardless of flux. At $\phi = 10^{17}$\,n\,cm$^{-2}$\,s$^{-1}$, a residence time $\tau_\mathrm{res} \lesssim 1$\,h with $\tau_\mathrm{hold} \sim 10$\,d would push $\mathcal{Y}_\mathrm{Au}^\mathrm{flow} > 0.999$ even at $\phi = 10^{18}$\,n\,cm$^{-2}$\,s$^{-1}$. This decouples the two timescales that create the flux optimum in the static model, allowing arbitrarily high flux without sacrificing gold yield. Combined with thermal-neutron treatment of \ce{^{196}Hg}, this enables nearly 100\% conversion.

\begin{figure}[tb]
\centering
\includegraphics[width=\columnwidth]{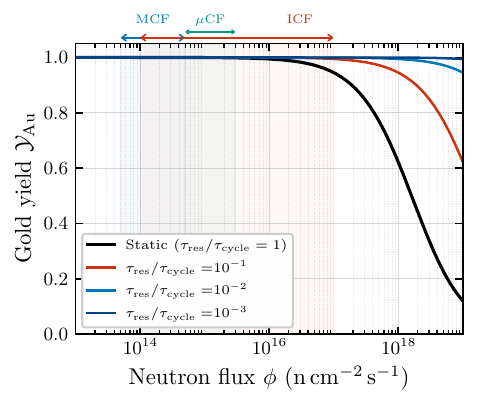}
\caption{Gold yield $\mathcal{Y}_\mathrm{Au}$ as a function of neutron flux for a static blanket ($\tau_\mathrm{res}/\tau_\mathrm{cycle} = 1$) and three flow-through duty cycles. The static blanket yield collapses above ${\sim}10^{15}$\,n\,cm$^{-2}$\,s$^{-1}$ due to gold-precursor burnup of \ce{^{197}Hg}. Flow-through operation recovers near-unity yield at arbitrarily high flux by limiting the time mercury spends in the neutron field.}
\label{fig:flow_through_yield}
\end{figure}

\begin{figure}[tb]
\centering
\includegraphics[width=\columnwidth]{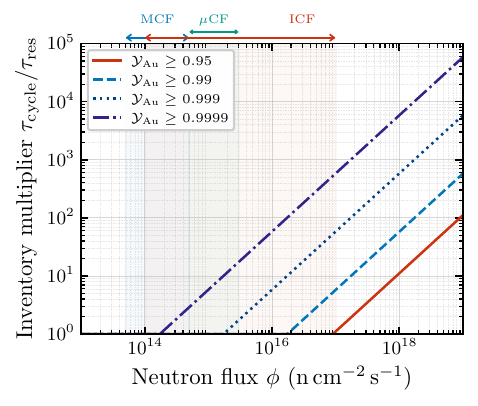}
\caption{Inventory multiplier $\tau_\mathrm{cycle}/\tau_\mathrm{res}$ required to maintain a target gold yield at each flux. At $\phi = 10^{17}$\,n\,cm$^{-2}$\,s$^{-1}$, achieving $\mathcal{Y}_\mathrm{Au} \geq 99\%$ requires holding ${\sim}100\times$ the in-flux inventory in the external hold-up volume. The multiplier grows as ${\sim}\sigma\phi/\lambda_\mathrm{g}$, i.e., linearly in flux, so the engineering cost scales mildly compared to the quadratic growth in transmutation rate.}
\label{fig:flow_through_inventory}
\end{figure}

Supplementary \Cref{fig:flow_through_yield} shows the gold yield as a function of flux for the static blanket and three flow-through duty cycles. The static blanket yield peaks near $\phi \sim 10^{15}$\,n\,cm$^{-2}$\,s$^{-1}$ and falls sharply at higher flux. With a duty cycle of $\tau_\mathrm{res}/\tau_\mathrm{cycle} = 10^{-2}$, the high-yield region extends to $\phi \sim 10^{17}$; at $10^{-3}$, it extends to $\phi \sim 10^{18}$ and beyond. Supplementary \Cref{fig:flow_through_inventory} quantifies the required inventory multiplier $\tau_\mathrm{cycle}/\tau_\mathrm{res}$: achieving $\mathcal{Y}_\mathrm{Au} \geq 99\%$ at $\phi = 10^{17}$\,n\,cm$^{-2}$\,s$^{-1}$ requires holding ${\sim}100\times$ the in-flux inventory in the external decay volume, a straightforward engineering constraint given that mercury is liquid at room temperature.

\section{Per-isotope value curves}
\label{supp:isotope_curves}

\begin{figure}[H]
\centering
\includegraphics[width=\columnwidth]{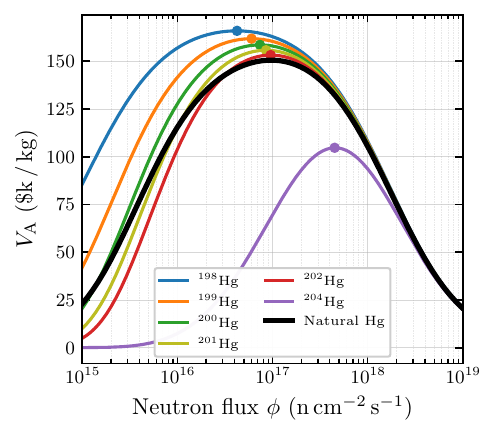}
\caption{Zoomed view of $V_\mathrm{A}(\phi)$ near the optimal-flux region ($r = 5\%$, $T \to \infty$). Dots mark the peak for each isotope. Longer chains peak at progressively higher flux following the $\sqrt{n}$ scaling. The solid black curve shows the abundance-weighted natural mercury value.}
\label{fig:isotope_value_zoomed}
\end{figure}

\begin{figure}[H]
\centering
\includegraphics[width=\columnwidth]{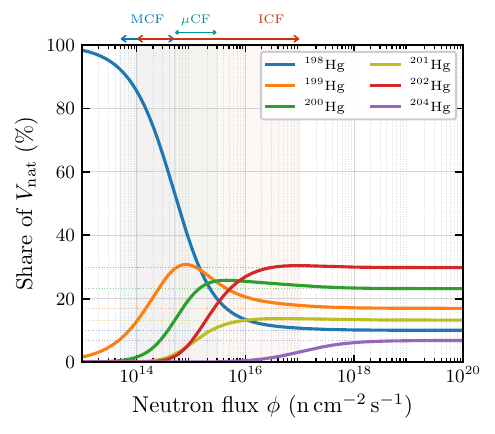}
\caption{Fractional contribution of each isotope to $V_\mathrm{nat}$ as a function of neutron flux ($r = 5\%$, $T \to \infty$). At low flux, \ce{^{198}Hg} contributes nearly all value; at $\phi \gtrsim 10^{17}$, the value shares converge toward the natural abundances (dotted horizontal lines).}
\label{fig:value_fraction}
\end{figure}

\section{Enrichment economics}
\label{supp:enrichment}

\subsection{Tails value and the cost of discarding heavy isotopes}

Let $E_\mathrm{F}$, $E_\mathrm{P}$, and $E_\mathrm{T}$ denote the \ce{^{198}Hg} mass fraction in the feed, product, and tails streams, respectively ($E_\mathrm{F} = 0.1002$ for natural mercury; $E_\mathrm{P} > E_\mathrm{F} > E_\mathrm{T}$). A feed mass $F$ is separated into product mass $P$ enriched to $E_\mathrm{P}$, and tails mass $T = F - P$ depleted to $E_\mathrm{T}$. The \ce{^{198}Hg} mass balance per kilogram of feed ($F = 1$) gives
\begin{equation}
    \frac{P}{F} = \frac{E_\mathrm{F} - E_\mathrm{T}}{E_\mathrm{P} - E_\mathrm{T}}, \qquad
    \frac{T}{F} = \frac{E_\mathrm{P} - E_\mathrm{F}}{E_\mathrm{P} - E_\mathrm{T}}.
    \label{eq:mass_balance}
\end{equation}
The NPV per kilogram of any mercury stream at enrichment fraction $E$ is
\begin{equation}
    V(E,\phi) = E\,V_{198}(\phi) + (1-E)\,\bar{V}_\mathrm{heavy}(\phi),
    \label{eq:VE}
\end{equation}
where $\bar{V}_\mathrm{heavy} \equiv \sum_{A>198} [f_A/(1-f_{198})]\,V_\mathrm{A}$ is the abundance-weighted average value of the heavy isotopes per kilogram of that fraction. This follows from equation~\eqref{eq:V_general} by setting $N_{198}(0) = E\,N_\mathrm{tot}$ and $N_j(0) = [f_j/(1{-}f_{198})]\,(1{-}E)\,N_\mathrm{tot}$ for $j > 198$. \Cref{eq:VE} assumes that the heavy-isotope fraction has the same relative composition in the product and tails streams (i.e., the process separates \ce{^{198}Hg} from the rest but does not sort among the heavy isotopes) Value is conserved across the cut for any $(E_\mathrm{P}, E_\mathrm{T})$,
\begin{equation}
\begin{aligned}
& \frac{P}{F}\,V(E_\mathrm{P}) + \frac{T}{F}\,V(E_\mathrm{T}) \\
& = \bigl[E_\mathrm{F}\,V_{198} + (1-E_\mathrm{F})\,\bar{V}_\mathrm{heavy}\bigr] = V_\mathrm{nat},
\end{aligned}
\label{eq:conservation}
\end{equation}
where the middle step follows directly from substituting Eqs.~\ref{eq:mass_balance}--\ref{eq:VE} and using $P E_\mathrm{P} + T E_\mathrm{T} = F E_\mathrm{F}$. Enrichment therefore never creates or destroys gold-production value; it only redistributes it between product and tails streams.

For the enrichment scenario of Table~2, the product is enriched to $E_\mathrm{P} = 90\%$ \ce{^{198}Hg}. Since the separation concentrates \ce{^{198}Hg} into the small product stream, the much larger tails stream is nearly depleted of \ce{^{198}Hg}: in the ideal limit $E_\mathrm{T} \to 0$, no \ce{^{198}Hg} leaks into the tails and they consist entirely of the heavy isotopes (\ce{^{199}}--\ce{^{204}Hg}). \Cref{eq:mass_balance} then gives $T/F = (E_\mathrm{P} - E_\mathrm{F})/E_\mathrm{P} = 0.799/0.90 = 0.888$\,kg of tails per kilogram of feed — most of the original mercury ends up as tails. The value per kilogram of tails is $V(E_\mathrm{T}{=}0) = \bar{V}_\mathrm{heavy}$. At $\phi = 10^{16}$\,n\,cm$^{-2}$\,s$^{-1}$ ($r = 5\%$, $T \to \infty$),
\begin{align}
    \bar{V}_\mathrm{heavy} &= \frac{1}{1-f_{198}}\sum_{A>198} f_A\,V_\mathrm{A} \notag\\
        &\approx 0.63\,p_\mathrm{Au} \approx \$111{,}000/\mathrm{kg},
    \label{eq:Vheavy_numeric}\\
    f_\mathrm{tails} \equiv \frac{(T/F)\,\bar{V}_\mathrm{heavy}}{V_\mathrm{nat}} &= \frac{0.888 \times 0.63}{0.660} \approx 85\%.
    \label{eq:tails_fraction}
\end{align}
Here $f_\mathrm{tails}$ is the fraction of the total feed NPV that resides in the tails stream; $1 - f_\mathrm{tails} \approx 15\%$ is captured by the product. Discarding the tails therefore destroys $f_\mathrm{tails}$ of the batch's gold-production value. This fraction grows with flux as the heavy-isotope values converge toward $V_{198}$ (Supplementary \cref{fig:value_fraction}), and falls toward zero at low flux where only \ce{^{198}Hg} is productive. The exact value depends weakly on $E_\mathrm{T}$ when $E_\mathrm{P}$ is close to $1$: even with $E_\mathrm{T} = 1\%$, the fraction remains ${\sim}85\%$.

\section{Blanket composition and facility economics}
\label{supp:blanket_economics}

\begin{figure*}[tb]
\centering
\includegraphics[width=1.6\columnwidth]{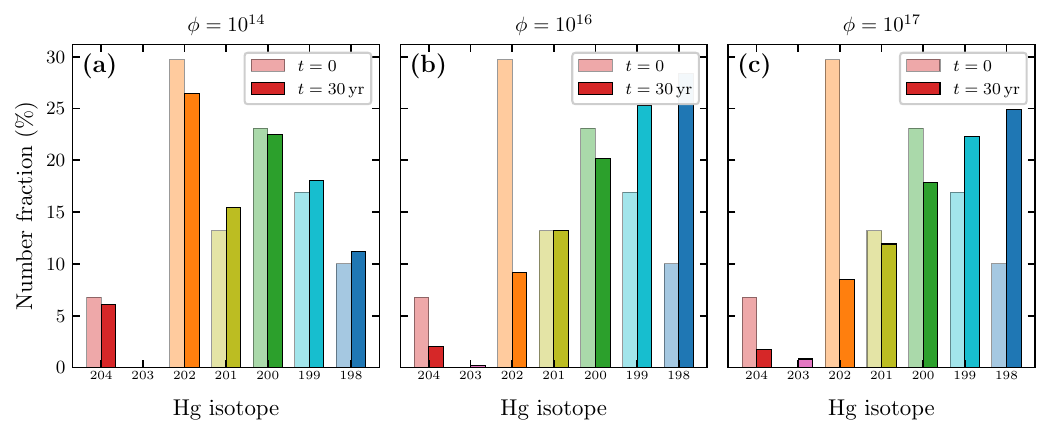}
\caption{Steady-state blanket isotope composition at $t = 0$ (faded) and $t = 30$\,yr (solid) for three neutron fluxes ($\sigma = 1.5$\,b, natural Hg with continuous replenishment). (a)~At $\phi = 10^{14}$\,n\,cm$^{-2}$\,s$^{-1}$, the composition changes only modestly. (b)~At $\phi = 10^{16}$\,n\,cm$^{-2}$\,s$^{-1}$, multi-step chains drive substantial in-situ enrichment: \ce{^{198}Hg} nearly triples from 10\% to 28\%. (c)~At $\phi = 10^{17}$\,n\,cm$^{-2}$\,s$^{-1}$, enrichment saturates with a visible \ce{^{203}Hg} inventory.}
\label{fig:composition}
\end{figure*}

\begin{figure*}[tb]
\centering
\includegraphics[width=1.4\columnwidth]{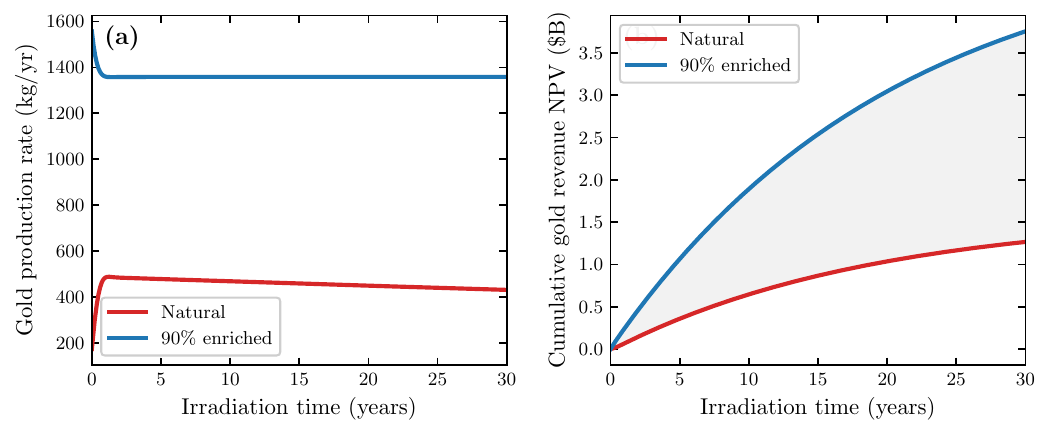}
\caption{Gold production from a 1\,GW D-T source at $\phi = 10^{17}$\,n\,cm$^{-2}$\,s$^{-1}$. (a)~Gold production rate vs.\ time for natural mercury (red) and 90\%-enriched \ce{^{198}Hg} (blue). (b)~Cumulative discounted gold revenue NPV ($r = 5\%$). The enriched facility reaches \$3.8\,B vs.\ \$1.3\,B for natural, a factor of ${\sim}3\times$ set by $\bar{n}_\mathrm{nat}/\bar{n}_\mathrm{enr} \approx 3.6/1.3$.}
\label{fig:facility_npv}
\end{figure*}

\section{Blanket time evolution}
\label{supp:time_evolution}

\begin{figure}[H]
\centering
\includegraphics[width=\columnwidth]{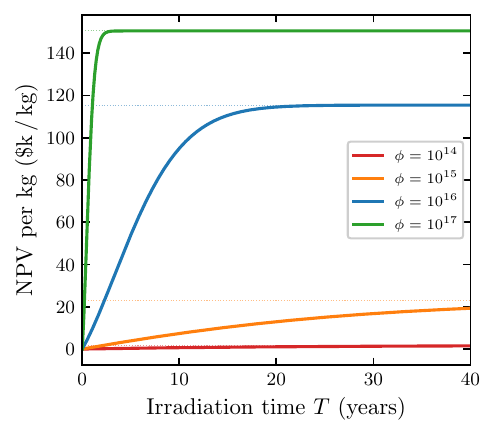}
\caption{Cumulative discounted gold revenue per kilogram of natural mercury blanket as a function of irradiation time at four neutron fluxes. Dots mark the half-maturity time $T_{50}$.}
\label{fig:maturity_npv}
\end{figure}

\begin{figure}[H]
\centering
\includegraphics[width=\columnwidth]{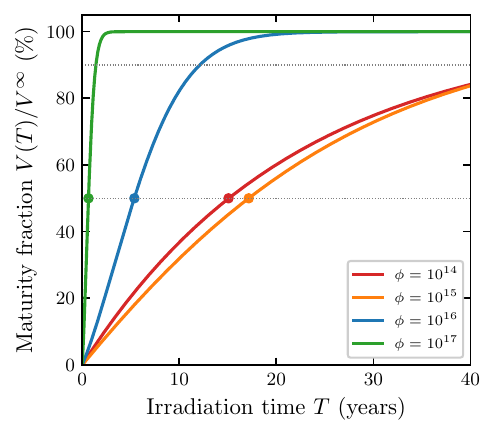}
\caption{Maturity fraction $V(T)/V^\infty$, with 50\% and 90\% guides. At $\phi = 10^{16}$\,n\,cm$^{-2}$\,s$^{-1}$, $T_{50} \approx 5$\,yr and $T_{90} \approx 11$\,yr.}
\label{fig:maturity_frac}
\end{figure}

\begin{figure*}[tb]
\centering
\includegraphics[width=1.4\columnwidth]{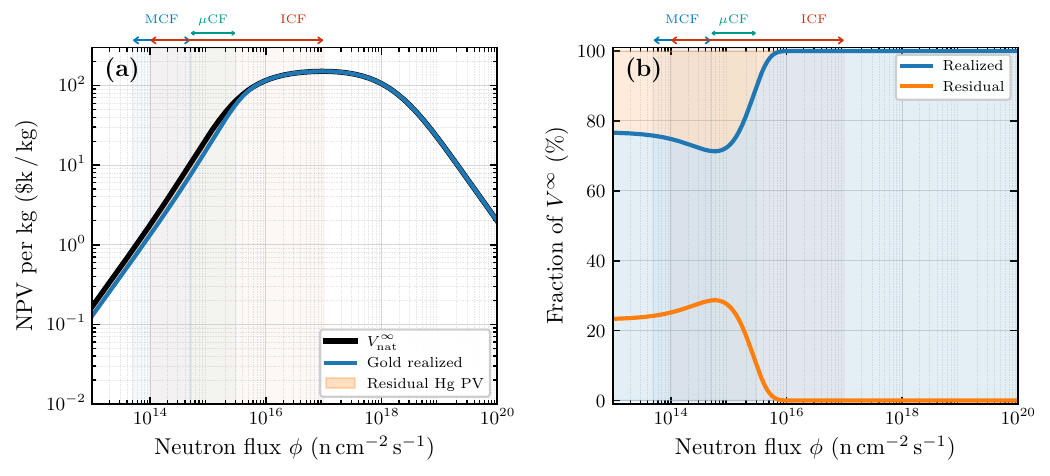}
\caption{Decomposition of $V_\mathrm{nat}^{\infty}$ into gold realized in $T = 30$\,yr and residual mercury present value ($r = 5\%$, $\sigma = 1.5$\,b, $b = 0.5$). (a)~Value per kilogram of natural mercury. (b)~Fraction of value realized as gold (blue) and retained as residual mercury (orange). Above $\phi \sim 10^{16}$, essentially all value is converted to gold. The residual fraction peaks near $\phi \sim 10^{15}$, where heavy-isotope chains first become valuable but cannot complete within 30\,yr.}
\label{fig:residual_value}
\end{figure*}

\begin{figure*}[tb]
\centering
\includegraphics[width=1.8\columnwidth]{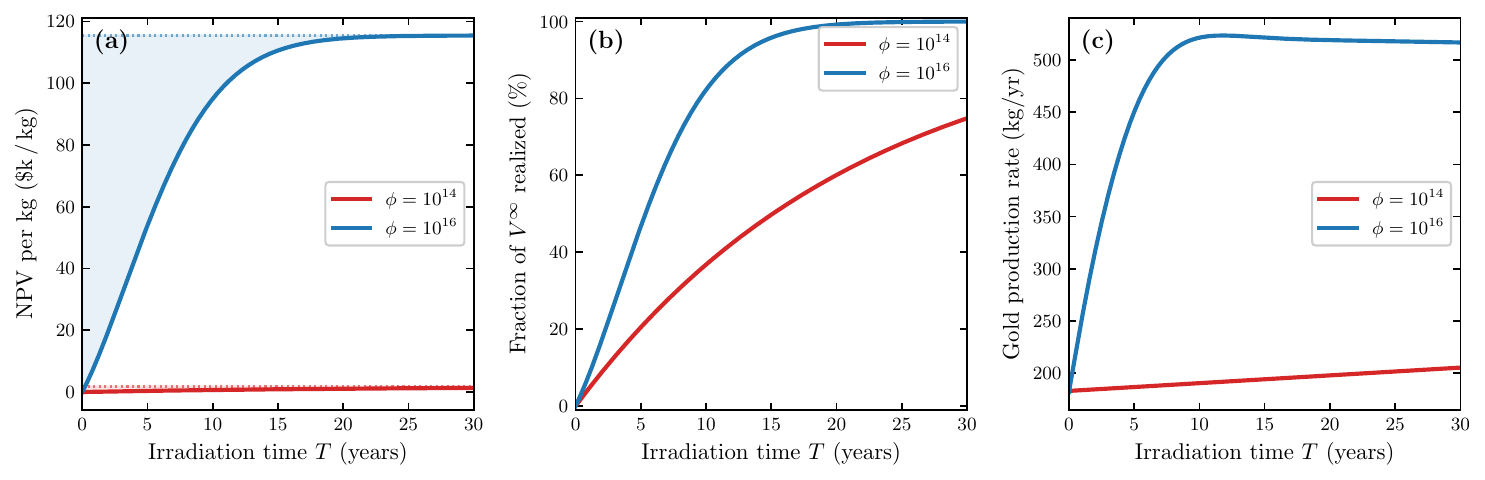}
\caption{Time-dependent profiles for two worked-example fluxes (1\,GW D-T source, $r = 5\%$, $\sigma = 1.5$\,b, $b = 0.5$, $\eta_\mathrm{pro} = 0.5$). (a)~Realized gold NPV per kilogram of natural mercury versus irradiation time; shaded regions show the residual mercury value. (b)~Fraction of $V^\infty$ realized as gold. (c)~Gold production rate in kg/yr from a continuously replenished natural mercury blanket.}
\label{fig:time_profiles}
\end{figure*}

\section{Conversion fraction and discount sensitivity}
\label{supp:conversion_sensitivity}

\end{document}